\documentclass[preprint2,tighten]{aastex62_reep}

\usepackage{epstopdf}
\usepackage{subfigure}  
\usepackage{amsmath}
\usepackage{hyperref}
\usepackage{graphbox}

\usepackage{longtable}

\usepackage{fancyhdr}   
\fancypagestyle{specialfooter}{%
 \fancyhf{}
 
 \fancyfoot[C]{Distribution Statement A.  Approved for public release: distribution is unlimited.}  }

\shorttitle{Mass Flows in Expanding Coronal Loops}
\shortauthors{Reep et al.}

\begin{document}

\title{Mass Flows in Expanding Coronal Loops}

\author[0000-0003-4739-1152]{Jeffrey W. Reep}
\affiliation{Institute for Astronomy, University of Hawai'i at M\=anoa, Pukalani, HI 96768}
\email{reep@hawaii.edu}

\author[0000-0001-8517-4920]{Roger B. Scott}
\affiliation{Space Science Division, Naval Research Laboratory, Washington, DC 20375}

\author[0000-0001-7754-0804]{Sherry Chhabra}
\affiliation{George Mason University, Fairfax, VA, 22030}

\author[0000-0002-7983-3851]{John Unverferth}
\affiliation{National Research Council Postdoc at the Naval Research Laboratory, Washington, DC 20375}

\author[0000-0002-2544-2927]{Kalman J. Knizhnik}
\affiliation{Space Science Division, Naval Research Laboratory, Washington, DC 20375}

\begin{abstract}
An expansion of cross-sectional area directly impacts the mass flow along a coronal loop, and significantly alters the radiative and hydrodynamic evolution of that loop as a result.  Previous studies have found that an area expansion from chromosphere to corona significantly lengthens the cooling time of the corona, and appears to suppress draining from the corona.  In this work, we examine the fluid dynamics to understand how the mass flow rate, the energy balance, and the cooling and draining timescales are affected by a non-uniform area.   We find that in loops with moderate or large expansion (cross-sectional area expansion factors of 2, 3, 10, 30, 100 from photosphere to apex), impulsive heating, for either direct thermal heating or electron beam heating, induces a steady flow into the corona, so that the coronal density continues to rise during the cooling phase, whereas a uniform loop drains during the cooling phase.  The induced upflow carries energy into the corona, balancing the losses from thermal conduction, and continues until thermal conduction weakens enough so that it can no longer support the radiative losses of the transition region (TR).  As a result, the plasma cools primarily radiatively until the onset of catastrophic collapse.  The speed and duration of the induced upflow both increase in proportion to the rate of area expansion.  We argue that observations of blue-shifted spectral lines, therefore, could place a constraint on a loop's area expansion.
\end{abstract}


\section{Introduction}

Models of closed coronal loops have typically assumed that the loops have a uniform cross-sectional area, although it is not clear whether this assumption is valid.  Naively, one might expect that loops must expand since the magnetic field falls off with height from the solar surface due to flux conservation, as in the solar wind (\textit{e.g.} \citealt{wang1995}).  However, observations of loop widths, which are used as a proxy for their cross-sections, have found that the expansion is quite limited along their lengths, significantly less than the change in magnetic field strength would imply \citep{klimchuk1992,klimchuk2000,klimchuk2020}.  \citet{peter2012} suggested that this may be because the temperature of the plasma is not constant transverse to the field line, while loops are typically observed in narrow temperature bands, causing an apparent constant width.  Direct measurements of the coronal magnetic field are likely required to resolve this contradiction between theory and observation.  

Assuming that the cross-sectional area of a loop does expand with height, as required by flux conservation, there must be a significant impact on the radiative and hydrodynamic evolution of that loop.  The mass flow rate, $\Dot{M} \equiv \rho A v$, is spatially constant in steady flow conditions, so both the mass density $\rho$ and bulk velocity $v$ are affected by changes in the area $A$.  The system of equations describing the long-term evolution of a loop is highly non-linear, and therefore this can drastically change the lifetime, emission in various wavelengths, and dynamic behavior of that loop.  \citet{cargill2022} showed that the expansion does in fact cause a large difference in (average) coronal densities and temperatures, while increasing the cooling time noticeably.  \citet{reep2022} similarly found that the densities, temperatures, and velocities are all significantly impacted in the presence of expansion, depending on the location and rate of expansion.  \citet{li1991} and the subsequent work by \citet{emslie1992} were the first to earnestly examine the impact of cross-sectional area on chromospheric evaporation.  They focused on the early heating phase, finding that the upflow speed is increased in the presence of expansion, and that the threshold for explosive evaporation depends weakly on the expansion factor ($\propto R_{m}^{1/5}$, where $R_{m}$, the magnetic mirror ratio, is their notation for expansion), in addition to the heating parameters \citep{fisher1985,reep2015}.

Curiously, \citet{reep2022} showed that it was immediately apparent that loops with a significant area expansion do not drain during their cooling phase, in contradiction to the well-known behavior of uniform area loops.  The commonly expressed $T \propto n^{2}$ scaling law relating the rate of cooling of the temperature $T$ to the rate of draining of the density $n$ of a coronal loop does not hold with area expansion.  In this paper, we focus on determining the cause of this unintuitive result.  We show that following impulsive heating, an excess of thermal conduction induces an upflow into the corona, lasting for significantly longer than the heating duration.  The loop is unable to drain while that upflow continues, so the density increases while the temperature decreases, unlike in a uniform loop where both decrease during the cooling phase.  

In this work, we examine the fluid dynamics to show the pertinent equations describing the velocity and Mach number for flows between the chromosphere and corona.  The draining timescale becomes effectively infinite prior to catastrophic cooling because the expansion induces a steady upflow of material into the corona during the radiative cooling phase.  We examine a set of numerical simulations to understand the role of expansion on evaporation and draining.  Finally, we synthesize a few spectral lines to understand how Doppler shifts might be impacted, and comment briefly on how that might be used to constrain whether and by how much the area expands.

\section{Fluid Dynamics}
\label{sec:fluid}

We wish to understand how flow speeds between the chromosphere and corona depend on the area expansion.  We first reiterate the derivation for a steady state flow of a compressible fluid in the presence of non-uniform cross-sectional area and heat transfer (\textit{e.g.} Section 6.8 of \citealt{saad1985}).  

Consider the set of equations given by

\begin{align}
    \partial_{t} \rho + \partial_{s} (\rho A v) &= 0  \\
    \partial_{t}(\rho v) + \frac{1}{A} \partial_{s}(A \rho v^{2}) + \partial_{s} P &= 0 \\
    P - \frac{\rho}{\mu} k_{B} T &= 0 \label{eqn:ideal} \\ 
    M \equiv \frac{v}{c_{s}} &= v \sqrt{\frac{\rho}{\gamma P}} \label{eqn:mach}
\end{align}

\noindent where the notation $\partial_{x}$ refers to the partial derivative with respect to variable $x$.  $\rho$ is the mass density, $A$ the cross-sectional area, $v$ the bulk flow velocity, $P$ the gas pressure, $T$ the temperature, $k_{B}$ the Boltzmann constant, $\mu$ the average particle mass, $c_{s}$ the sound speed, and $\gamma$ the ratio of specific heats.  These are the equations for the conservation of mass and momentum in the direction of flow $s$, the equation of state for an ideal gas, and the definition of Mach number (or sound speed).   We have neglected the effects of gravity and viscosity in the momentum equation, and we have omitted the energy equation in favor of an assumed temperature profile $T(s)$.  From these equations, we can derive a relation between the change in Mach number $M$ in terms of the change in area $A$ and change in temperature $T$, to better understand flows in an expanding loop.  

We first assume a steady state, that is, $\partial_{t} = 0$.  We can then write the continuity equation as:

\begin{align}
    \partial_{s} (\ln{A}) + \partial_{s}(\ln{\rho}) + \partial_{s}(\ln{v}) = 0 \label{eqn:cont}
\end{align}

\noindent and the momentum equation:

\begin{align}
\partial_{s} P + \rho v^{2} \partial_{s}(\ln{A}) + v^{2} \partial_{s} \rho + 2 \rho v \partial_{s} v &= 0 \nonumber \\
\partial_{s} P + \rho v \partial_{s} v &= 0  
\label{eqn:mom}
\end{align}

\noindent where we have used the steady-state continuity equation to simplify the result.

We next derive a relation between the change in pressure, density, and temperature by differentiating the equation of state:

\begin{align}
    \partial_{s} (\ln{P}) = \partial_{s} (\ln{\rho}) + \partial_{s} (\ln{T})
\end{align}

\noindent and then similarly for the definition of the Mach number:

\begin{align}
    \partial_{s} (\ln{v}) = \partial_{s} (\ln{M}) + \frac{1}{2} \partial_{s} (\ln{T}) \label{eqn:mach2}
\end{align}

Equations \ref{eqn:mach} through \ref{eqn:mach2} can then be combined to find the following relation:

\begin{align}
    (1 - \gamma M^{2}) \partial_{s} (\ln{M}) = & \Big(\frac{\gamma M^{2} + 1}{2}\Big) \partial_{s} (\ln{T}) \notag \\ & - \partial_{s} (\ln{A})
    \label{eqn:full}
\end{align}

\noindent This is Equation 6.81 of \citet{saad1985}.  This relation generally shows that the change in Mach number is proportional both to the change in area and the change in temperature.  

A simplified form of this equation is commonly used, but does not apply to the chromosphere-corona boundary.  In the absence of heat flow, the isentropic limit of the equation can be derived from Equations \ref{eqn:cont}, \ref{eqn:mom}, and the definition of sound speed at constant entropy, $c_{S} = \sqrt{(\frac{\partial P}{\partial \rho})_{S}}$ to give:

\begin{align}
    \partial_{s} (\ln{v}) = \frac{1}{M_{S}^{2} - 1} \partial_{s} (\ln{A}),
\end{align}

where $M_{S} = v / c_{S}$.
In the adiabatic limit, therefore, the change in flow speed $v$ is directly related to the change in area $A$.  For subsonic flows, the speed increases with decreasing area and decreases with increasing area, and vice versa for supersonic flows.  \citet{emslie1992} cited the isentropic limit to examine chromospheric evaporation due to heating from an electron beam, where material upflows from chromosphere to corona.  However, the temperature jumps 2 to 3 orders of magnitude between those heights, so this limit does not generally apply.  

Equation \ref{eqn:full} represents the full relation between the area, Mach number, and temperature.  Importantly, it does not depend on the specifics of what causes the change in temperature or area, so we do not need to specify the exact forms of heating or cooling by radiation or conduction.  In this work, we are primarily concerned with flows between the chromosphere and corona, which are generally subsonic, and so this equation can be further simplified.

\subsection{Subsonic Limit}

In the subsonic limit, $M < 1$, Equation \ref{eqn:full} can be simplified:

\begin{align}
    \partial_{s} (\ln{M}) = \frac{1}{2} \partial_{s} (\ln{T}) - \partial_{s} (\ln{A}) \label{eqn:dM}
\end{align}

\noindent We can then integrate this from chromosphere (1) to corona (2) to obtain:

\begin{align}
    \ln{\frac{M_{2}}{M_{1}}} &= \frac{1}{2} \ln{\frac{T_{2}}{T_{1}}} - \ln{\frac{A_{2}}{A_{1}}} \nonumber \\
    M_{2} &= M_{1} \Big(\frac{T_{2}}{T_{1}} \Big)^{\frac{1}{2}} \frac{A_{1}}{A_{2}} 
\end{align}

\noindent For the corona-chromosphere transition, $\frac{T_{2}}{T_{1}} \gtrsim 100$, so the Mach number should increase across the TR and into the corona.  With a uniform area, the heat flow therefore dominates this change.  For an increasing area, $A_{2} > A_{1}$, however, the Mach number in the corona can be reduced significantly.  

We can carry this argument further.  Using Equations \ref{eqn:ideal} and \ref{eqn:mach}, we can rewrite Equation \ref{eqn:dM} in terms of the flow speed:

\begin{align}
    \partial_{s} (\ln{v}) &= \partial_{s} (\ln{T}) - \partial_{s} (\ln{A}) 
\end{align}

\noindent and again integrating from chromosphere (1) to corona (2)

\begin{align}
    \frac{v_{2}}{v_{1}} &= \frac{T_{2}}{T_{1}} \frac{A_{1}}{A_{2}}  \label{eqn:vel}
\end{align}

\noindent The velocity in the corona is reduced due to an area expansion, and increased for an area contraction (as with the isentropic limit).  Note, however, that the dependence on temperature is linear.  A factor of 10 increase in the temperature would increase the velocity by 10, but the Mach number only by about 3.  

Furthermore, combining Equation \ref{eqn:vel} with the continuity equation and ignoring gravity, we can also write 

\begin{align}
    \frac{T_{2}}{T_{1}} = \frac{\rho_{1}}{\rho_{2}}
\end{align}

\noindent which is equivalent to a condition of constant pressure.  The density profile changes inversely with the temperature profile, but is otherwise unaffected by any change in the cross-sectional area.  

Finally, we can combine the expressions for velocity and Mach number to derive the change in sound speed:

\begin{align}
    \frac{c_{2}}{c_{1}} = \Big(\frac{T_{2}}{T_{1}} \Big)^{\frac{1}{2}}
\end{align}

\noindent which also follows directly from the definition of the sound speed ($c_{s} \propto \sqrt{\frac{P}{\rho}} \propto \sqrt{T}$).

Assuming that the temperature jumps from 10$^{4}$ to 10$^{6}$ K between the chromosphere and corona ($\frac{T_{2}}{T_{1}} \approx 100$), the relations we have are thus:

\begin{align}
   \frac{v_{2}}{v_{1}} &\approx 100 \frac{A_{1}}{A_{2}} \\
   \frac{M_{2}}{M_{1}} &\approx 10 \frac{A_{1}}{A_{2}} \\
   \frac{c_{2}}{c_{1}} &\approx 10
\end{align}

\noindent For a constant area expansion, $A_{1} = A_{2}$, this implies that the velocity in the corona is a factor of 100 times that in the chromosphere, while the Mach number and sound speed are both only 10 times higher.  However, consider an expansion by a factor of 10, that is $A_{2} = 10 A_{1}$.  In this case, the Mach number is approximately constant from chromosphere to corona, and the velocity is only 10 times higher in the corona.  The flows effectively become slowed in the corona due to the area expansion.  This is apparent from the simulations in \citet{reep2022} and the simulations in Section \ref{sec:modeling} of this work.

\subsection{Sustained Upflows during the Cooling Phase of Impulsively-Heated Expanding Loops}

During the cooling phase of an impulsively heated coronal loop, there is a relation between temperature and density, $T \propto n^{\delta}$, where $\delta$ is a parameter that relates how quickly the loop cools versus drains.  Early hydrodynamic modeling found $\delta \approx 2$ during the radiative cooling phase \citep{serio1991,jakimiec1992}, while later results showed that the value of $\delta$ depends on the shape of the radiative loss function \citep{cargill1995}.  \citet{bradshaw2005,bradshaw2010} derived a full expression: $\delta = \gamma - 1 + \frac{\tau_{v}}{\tau_{R}}$, where $\gamma$ is the ratio of specific heats, $\tau_{v}$ is the draining timescale of the loop, and $\tau_{R}$ is the radiative timescale.  The draining rate is intimately tied to the cooling by radiation of a coronal loop.  For adiabatic cooling, $\tau_{v} \rightarrow 0$, so $\delta = \gamma - 1$, while for purely radiative cooling $\tau_{v} \rightarrow \infty$, or $\delta \rightarrow \infty$.    

All of these initial results were derived and simulated, however, assuming that the coronal loop had a uniform area across its length.  Recent work has begun to examine the consequences of ignoring this assumption.  \citet{cargill2022} noted that even modest area expansion affects both thermal and enthalpy fluxes between the corona and TR, and can lead to radically different hydrodynamic evolution.  \citet{reep2022} showed that for much larger area expansions, the loop does not drain significantly during the cooling phase, implying that the cooling occurs purely radiatively, or that $\delta \rightarrow \infty$.  In fact, with a large area expansion, there is an induced upflow into the corona that lasts for significantly longer than the heating duration (see Figure 2 of \citealt{reep2022}), supplying mass and energy to the corona until it ceases.  The loop is unable to drain while there is an upflow of material from the chromosphere, so it is in a regime where the coronal density is rising marginally while the temperature is falling due to radiation.  

Why, then, does material continue to evaporate from the chromosphere after heating ceases?  
The conductive heat flux vanishes at the base of the TR, and as long as the upflow persists the enthalpy flux is directed into the corona, so the only term removing removing energy from the loop is radiation.  The timescale for the loop to cool is, therefore, bounded from below by the total thermal energy in the loop divided by the integrated radiative losses over the whole loop.  This cooling time is longer than the timescale of any individual term in the energy equation, meaning that the plasma in the loop is in approximate equilibrium.  Neglecting gravity and viscosity, the energy balance is approximately given by:

\begin{align}
    \frac{1}{A} \partial_{s} \Big(A v (P+E )\Big) &- \frac{1}{A} \partial_{s} \Big(\kappa_{0} A T^{5/2} \partial_{s} T \Big)  \nonumber \\
    &+ n_{e} n_{H} \Lambda(T, Z) = 0
\end{align}

\noindent where the three terms are, respectively, the enthalpy flux, thermal conduction, and radiative losses.  $P$ is the pressure, $E$ the internal energy, $\kappa_{0}$ the coefficient of thermal conductivity, $n_{e}$ and $n_{H}$ the electron and hydrogen densities, and $\Lambda(T, Z)$ the radiative loss function, which depends on the temperature $T$ and abundances $Z$.  Using the chain rule, we can expand this as:

\begin{align}
    \partial_{s} \Big(v (P+E )\Big) - \partial_{s} \Big(\kappa_{0} T^{5/2} \partial_{s} T \Big) + n_{e} n_{H} \Lambda(T, Z) \nonumber \\
    + \Big[v(P+E) - \kappa_{0} T^{5/2} \partial_{s} T \Big] \partial_{s} (\ln{A} ) = 0
    \label{eqn:new_balance}
\end{align}

\noindent The first three terms are the energy balance of a loop with uniform cross-sectional area.  However, the presence of non-uniform area introduces an extra term for both thermal conduction and enthalpy flux that depend on the rate of change of the area along the loop length.  Additionally, since conduction flows from high temperature to low temperature (corona to chromosphere), while the mass flows from high pressure to low pressure (chromosphere to corona), these two terms are oppositely directed.

As plasma evaporates into the loop, raising the density significantly, the initially small conductive timescale $\tau_{C} \propto \frac{n L^{2}}{T^{2.5}}$ grows, while the initially large radiative timescale $\tau_{R} \propto \frac{T^{1.5}}{n}$ falls.  The cooling eventually becomes driven primarily by radiation, albeit weakly at first and growing with time.  While that radiation is weak, we can see that if the area expansion is sufficiently large, so that $\partial_{s} (\ln{A})$ is large, the only way for Equation \ref{eqn:new_balance} to hold is for the term multiplying $\partial_s \ln A$ to vanish, i.e.,

\begin{align}
 \Big[v(P+E) - \kappa_{0} T^{5/2} \partial_{s} T \Big]\approx 0,
 \label{eqn:parity}
\end{align}

\noindent so the energy carried by flows into the corona approximately balances the energy lost due to conduction.  That is, these two terms cause no net change in the energy of the system, but the flows into the corona must be maintained to balance the conductive cooling, which continues as long as the radiative loss term is relatively small in the energy balance equation. 

This balance between the heat flux and the enthalpy flux does not hold in the TR, where the enthalpy flux (which scales with the sound speed) becomes small and radiation becomes large.
However, because the mass flux is conserved along the lower portion of the loop, the upflow implied by the coronal enthalpy flux extends all the way to the chromosphere.  The induced upflow is effectively a coronal heating term that balances conductive cooling.  The increased cooling time in expanded loops is thus due to energy loss being only from radiation, rather than a combination of radiation and enthalpy losses (as in \citealt{bradshaw2010}).  Additionally, for a heating event of equal magnitude, the coronal density is lower in loops with larger expansion (see \citealt{cargill2022}, \citealt{reep2022}, or Figure \ref{fig:heated}), so the radiative timescale is somewhat larger with expansion.  

Since the radiative loss function $\Lambda(T,Z)$ grows as the temperature falls and since the density is being consistently raised by upflows, the radiation term, $n_{e}n_{H} \Lambda(T,Z)$, grows large during the cooling phase.  At the TR, where the conductive flux supplies the heat necessary to maintain the radiation, it eventually grows large enough that conduction cannot maintain it.  When this occurs, the upflows cease, and the corona begins to drain, so that the enthalpy flux begins to then carry energy out of the corona (becomes a cooling term).  The conductive flux continues to carry energy out of the corona, as well, so the approximate parity in the flux terms in Equation \ref{eqn:parity} is then broken.  This marks the onset of catastrophic cooling and the loop collapses in short order.  
 
%Compared to the uniform case, an area expansion from chromosphere to corona increases the energy delivered by thermal conduction to the base of the loop, as shown in Equation \ref{eqn:new_balance} (see also \citealt{mikic2013}).  This in turn raises the pressure at the top of the chromosphere, driving an expansion of material back into the corona.  The induced flow carries both mass and energy into the corona, so the enthalpy flux is effectively a heating term. The cooling time of the corona is therefore increased, and, since there is a net flow of material into the corona, the density can only increase. The corona therefore is primarily cooled through radiation, which gradually carries energy out of the system, lowering the temperature of the corona.  As the coronal temperature falls, the radiative loss rate $\Lambda(T)$ increases (peaking near $10^{5}$ K), while thermal conduction weakens.  Once the temperature has decreased enough that the conductive flux is weaker than the radiative losses in the lower atmosphere, the pressure imbalance dissipates and the upflow ceases.  Since the magnitude of the additional conduction term $\propto \partial_{s} (\ln{A})$, the duration of that induced flow depends on how much and how quickly the area expands, which we will show. 

This suggests that with an area expansion, the evolution of an impulsively heated loop proceeds in a different regime from the usual paradigm.  In uniform area loops, the cooling is initially dominated by thermal conduction, gradually transitioning to radiative and enthalpy-driven cooling, before a catastrophic collapse \citep{cargill1995,bradshaw2005,bradshaw2010}.  During this cooling, however, a uniform loop also drains significantly, leading to the $T \propto n^{\delta}$ scaling.  Conversely in an expanding loop, the induced upflow prevents draining as flows carry mass and energy into the corona, while the loop slowly cools through radiation.  The result is that the cooling time is lengthened in proportion to $\ln{(A)}$, while draining begins at the onset of catastrophic collapse.  This was shown in \citet{reep2022}, and will be explored further in the next section.

\section{Modeling}
\label{sec:modeling}

We now use hydrodynamic modeling to explore the behavior of flows in loops with area expansion.  In the derivations in Section \ref{sec:fluid}, we have made simplifying assumptions that the flows are approximately in steady state, and we have neglected the effects of gravity and viscosity.  To get away from those assumptions, we use a model that treats the physics more fully to understand the behavior of the system.

We have run the simulations with the open source HYDrodynamics and RADiation code (HYDRAD\footnote{https://github.com/rice-solar-physics/HYDRAD}; \citealt{bradshaw2013}).  The code solves the Navier-Stokes equations for plasma constrained to move along a magnetic flux tube, including the effects of variable cross-sectional area.  The full equations with area expansion were detailed in \citet{reep2022}.  The code, written in C++, is robust, lightweight, and capable of being run on a desktop computer.  The radiative losses have been calculated with the CHIANTI atomic database \citep{dere1997}, version 10 \citep{delzanna2021}, assuming equilibrium ionization populations.

\subsection{Impulsive heating}
\label{subsec:impulsive}

\begin{figure*}
    \centering
    \includegraphics[width=0.45\textwidth]{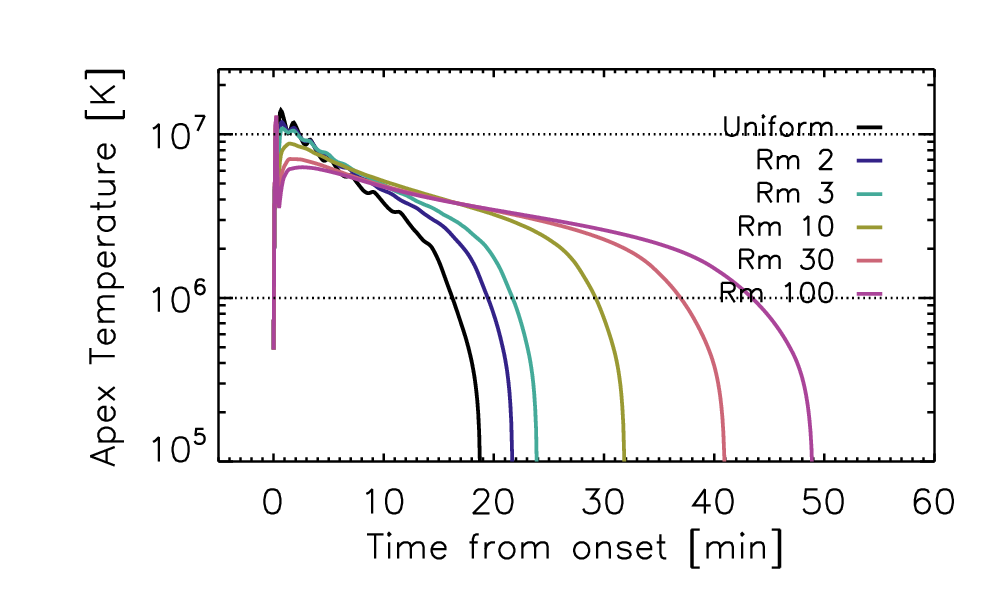}
    \includegraphics[width=0.45\textwidth]{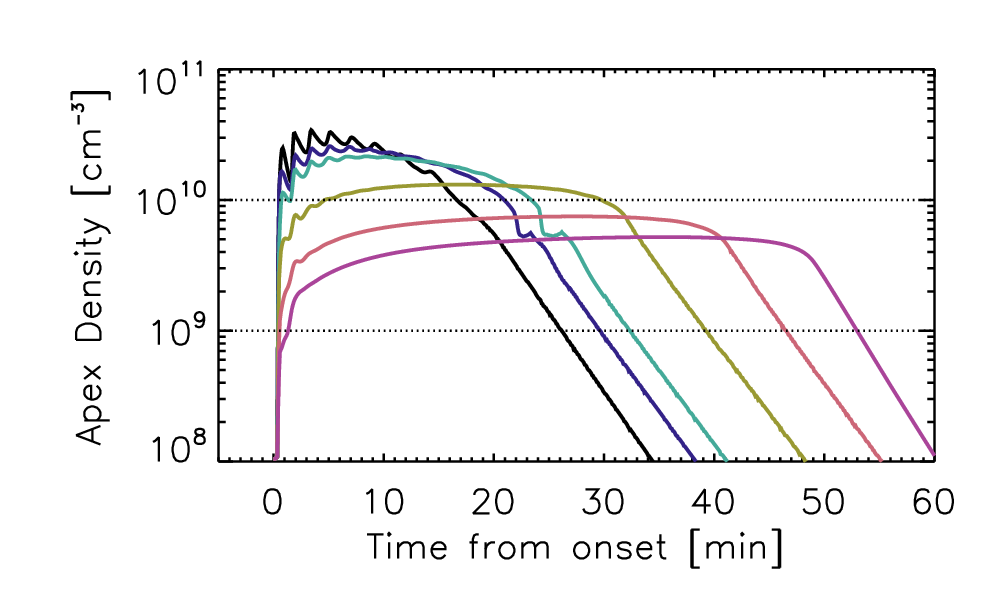}
    \includegraphics[width=0.45\textwidth]{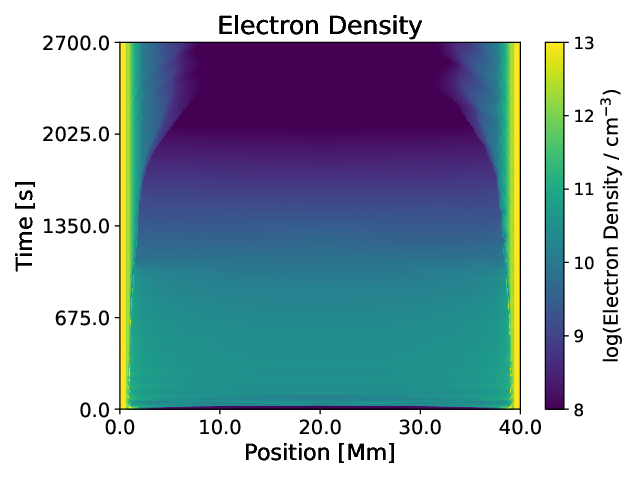}
    \includegraphics[width=0.45\textwidth]{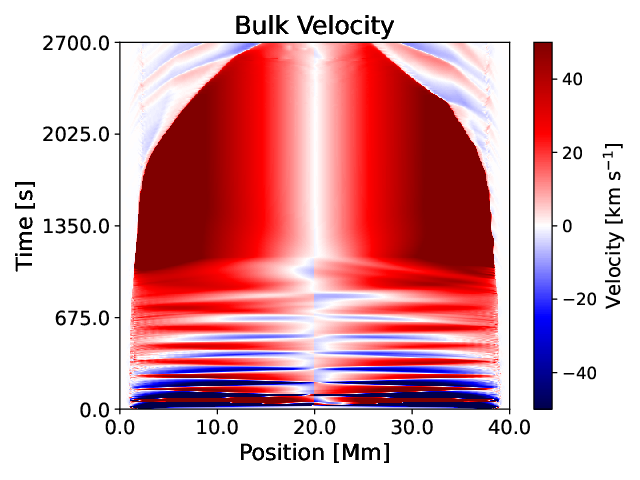}
    \includegraphics[width=0.45\textwidth]{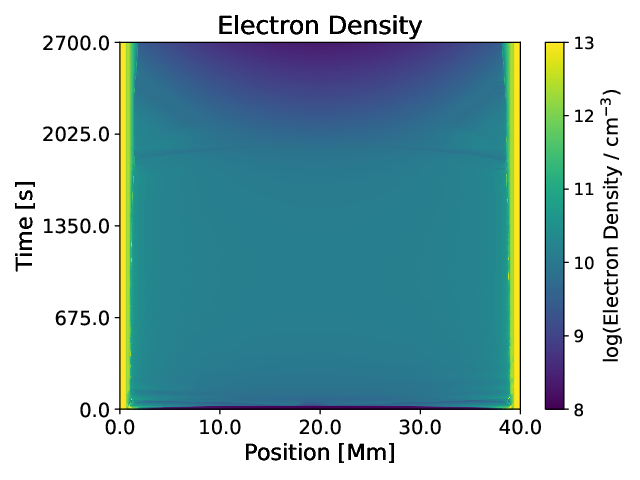}
    \includegraphics[width=0.45\textwidth]{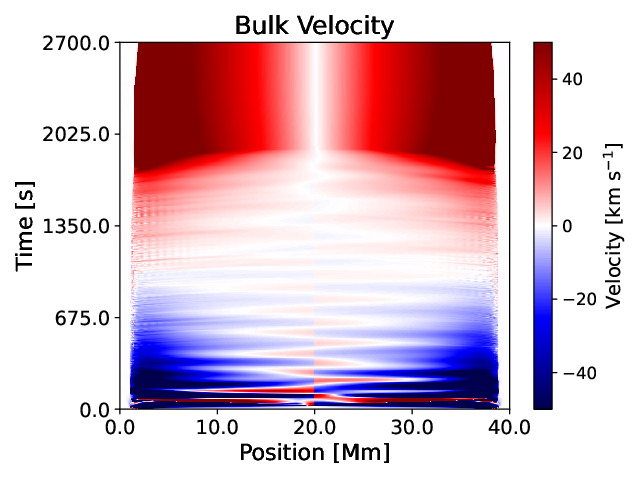}
    \caption{A comparison of the evolution of loops heated impulsively with an electron beam for 20 s.  The top row shows the apex temperatures and densities of four 40 Mm loops, with expansion factors of 1, 10, 30, and 100 from footpoint to apex.  The middle and bottom rows show the spatiotemporal evolution of the electron densities and bulk velocities for the cases with uniform area (middle) and expansion of 10 (bottom).  With area expansion, there is an induced upflow that lasts for significantly longer than the heating phase, that gradually raises the density of the loop during the radiative cooling phase.  Note that the color bar on the velocity plots saturate at $\pm 50$ km s$^{-1}$, though the speeds reach higher values.\label{fig:heated}}
\end{figure*}

\begin{figure*}
    \centering
    \includegraphics[width=0.49\textwidth]{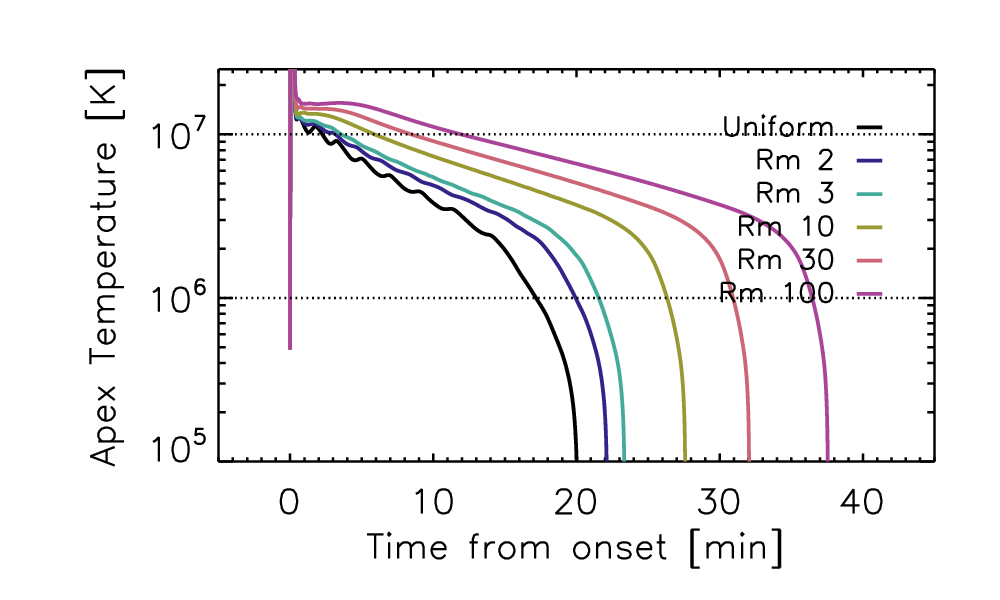}
    \includegraphics[width=0.49\textwidth]{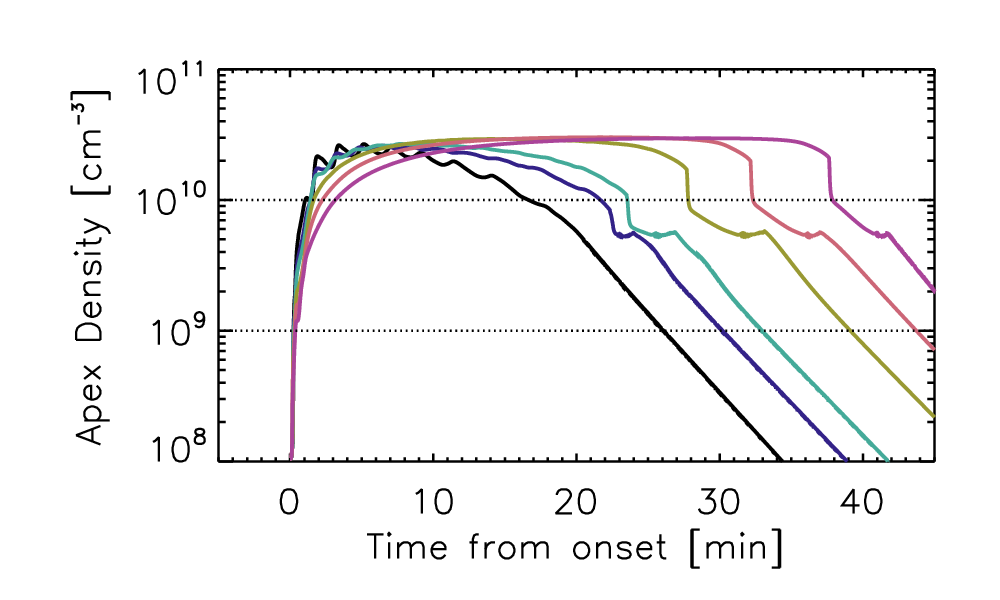}
    \includegraphics[width=0.49\textwidth]{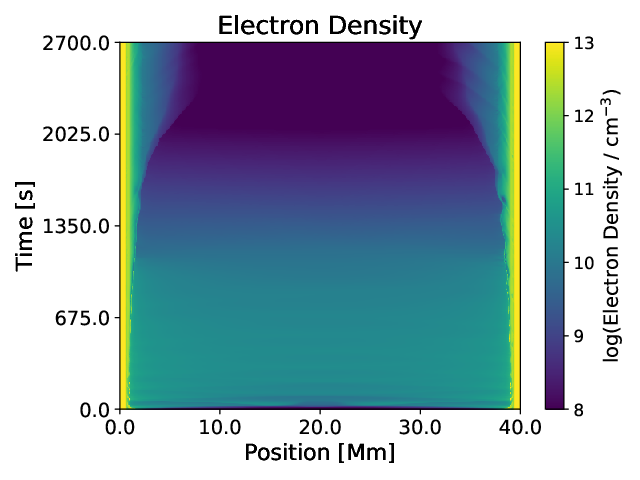}
    \includegraphics[width=0.49\textwidth]{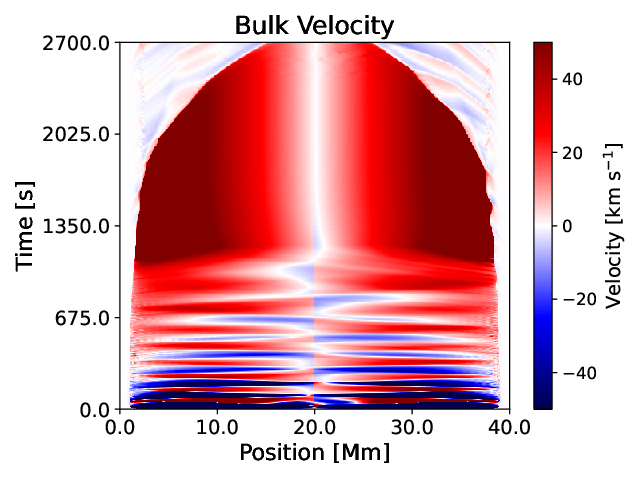}
    \includegraphics[width=0.49\textwidth]{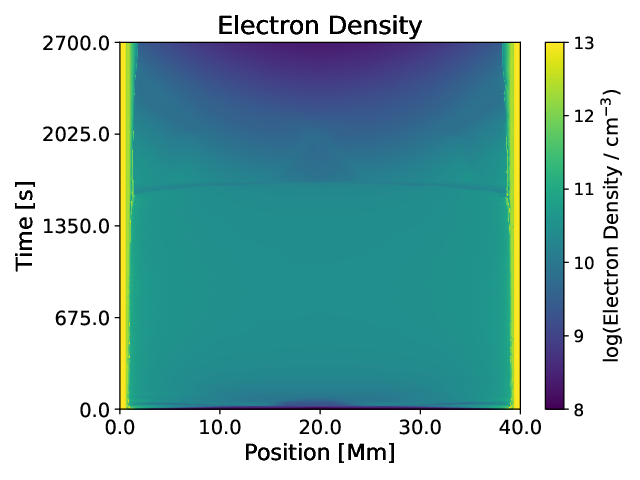}
    \includegraphics[width=0.49\textwidth]{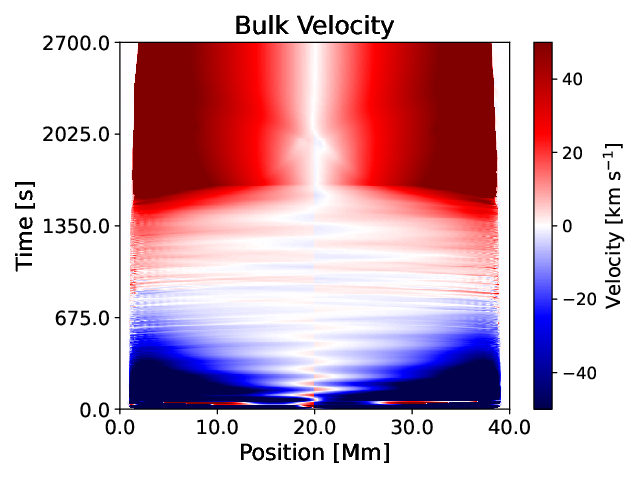}
    \caption{A comparison of the evolution of loops heated impulsively with a thermal pulse for 20 s.  Similar to Figure \ref{fig:heated}.  The densities and temperatures reach higher values in this case, but in loops with expansion there is again a long-lasting upflow during the cooling period that prevents draining.  That is, the flow behavior does not depend on the heating mechanism.  \label{fig:thermal}}
\end{figure*}

We first examine the behavior of loops heated with an impulsive burst of energy.  We show a brief comparison of the evolution of loops of varying area expansion, with two sets of simulations showing heating with an electron beam (Figure \ref{fig:heated}) and with a purely thermal impulse (Figure \ref{fig:thermal}).  We assume that the loops have area expansion factors of 1 (uniform), 2, 3, 10, 30, and 100 from footpoint to apex, using a functional form of $A(s) = 1 + A_{0} \sin^{2}{(\frac{\pi s}{2L})}$, for $s$ the curvilinear coordinate and $2L$ the total loop length.  We choose values of $A_{0} = [0, 1, 2, 9, 29, 99]$ to give the maximal areas, where the magnitudes are based on the range of values inferred from magnetic extrapolations (see the appendix of \citealt{reep2022}).  We use as a shorthand the notation $R_{m}$ (mirror ratio) from \citet{emslie1992} to denote the ratio of footpoint-to-apex expansion (occasionally $\Gamma$ has also been used in the literature).

In Figure \ref{fig:heated}, we impulsively heat these loops with a moderate electron beam for 20 s on a triangular profile, peaking at $5 \times 10^{10}$ erg s$^{-1}$ cm$^{-2}$, which raises the coronal temperature to around 10 MK and drives evaporative flows into the corona.  The top row shows the comparison of the evolution of the apex temperatures (left) and densities (right).  The bottom two rows show a comparison of the spatiotemporal variation of the electron density $n_{e}$ (left) and bulk flow velocity $v$ (right) for the uniform loop (middle row) and expansion of 10 (bottom row).  The x-axis shows position along the loop, while the y-axis shows time evolution.  Blue flows are towards the apex, while red are away from the apex.  

We can see that the cooling times are significantly lengthened with area expansion.  Additionally, with expansion, we see that there is an upflow that lasts significantly longer than the heating period (which lasted for 20 s).  The duration of this upflow, moreover, is directly related to the expansion factor.  After 200 s, the evaporation has essentially ceased in the uniform case, has weakened with an expansion of 10, and is still strong with an expansion of 100.  With expansion, the coronal density continues to increase, rather than drain, during the radiatively cooling phase, in direct contrast to the gradual draining that occurs in a uniform-area loop.  These results can be compared to \citet{reep2022}, which examined the dynamics and resultant radiative output more directly.  

In Figure \ref{fig:thermal}, we show a similar comparison of six loops heated with a thermal pulse.  We assume the same total energy injection rate as the beam heating case, for 20 s of total heating, but in this case, the heating is distributed uniformly over the length of the loop.  As a result, the temperatures and densities reach higher values than the beam-heated case.  The long-lasting upflow is similarly present for expanding loops in this case, showing that the qualitative behavior does not depend directly on the assumed heating mechanism (which also could've been intuited from the results of \citealt{cargill2022}).  We therefore expect similar flow behavior ought to occur in both quiescent coronal loops and flaring loops.

\begin{figure*}
    \centering
    \includegraphics[width=0.49\textwidth]{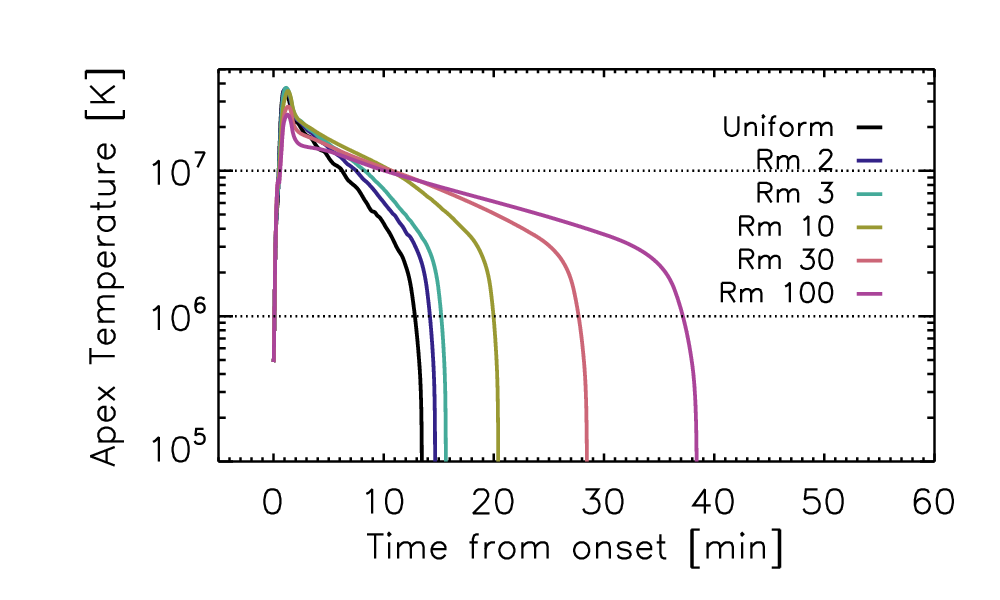}
    \includegraphics[width=0.49\textwidth]{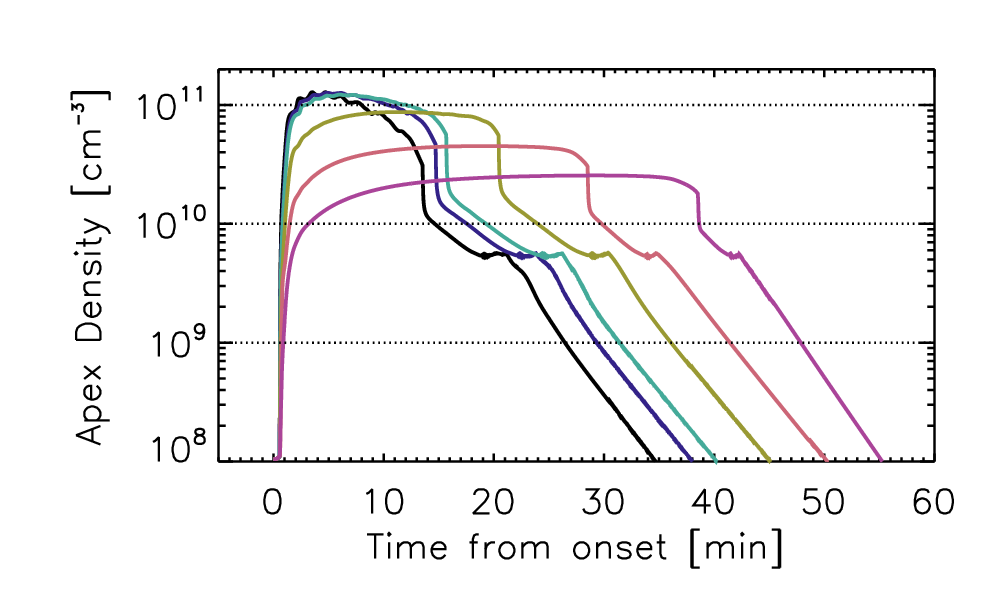}
    \includegraphics[width=0.49\textwidth]{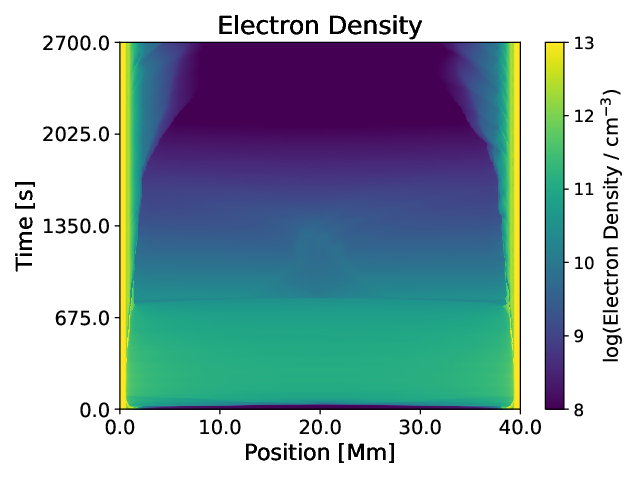}
    \includegraphics[width=0.49\textwidth]{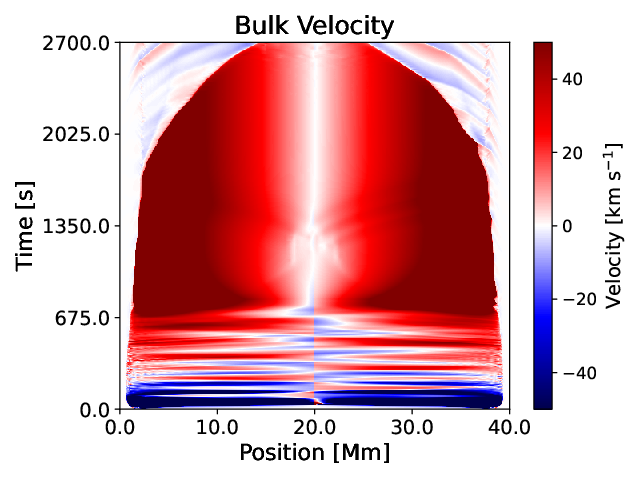}
    \includegraphics[width=0.49\textwidth]{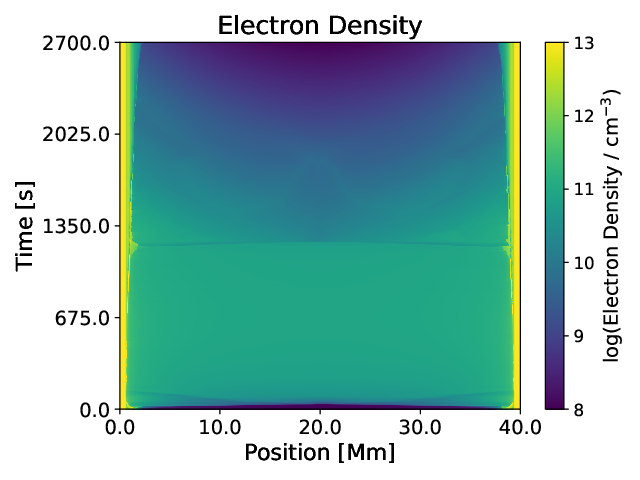}
    \includegraphics[width=0.49\textwidth]{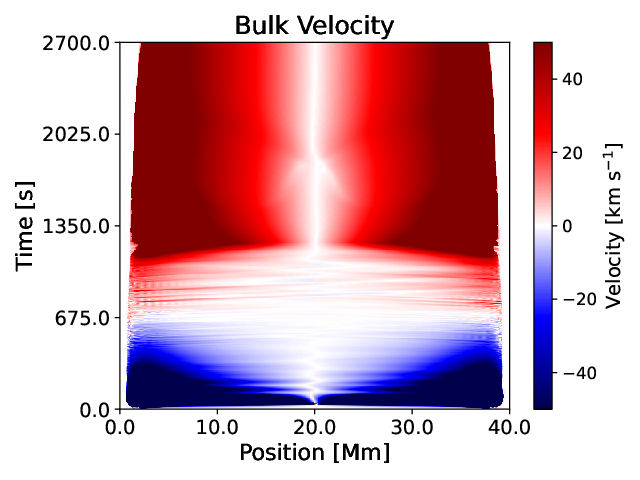}
    \caption{A comparison of the evolution of loops heated impulsively with a beam for 100 s.  Similar to Figure \ref{fig:heated}.  The induced upflow is stronger, but of similar duration, to that induced by the shorter duration beam.  The loops reach higher temperatures and densities, and as a result collapse more quickly.  \label{fig:strongbeam}}
\end{figure*}

Finally, as one more point of comparison, we include a case where we heat the set of loops with a beam lasting for a total of 100 s duration, with a triangular profile ramping up over 50 s and down over 50 s.  Figure \ref{fig:strongbeam} shows this comparison, with plots similar to the previous two figures.  In this case, the induced upflow reaches higher speeds, but lasts for a similar amount of time.  As a result, the plasma reaches higher temperatures and densities than in Figure \ref{fig:heated}, and so the loops cool more quickly.  In the case of uniform loops, the heating duration is directly tied to the upflow duration \citep{reep2018}, whereas here it appears that expansion is the more important factor.

\subsection{Steady-state Flows}

The derivations in Section \ref{sec:fluid} assumed that the flows were steady state.  When we examine the flows between the corona and chromosphere in the simulations, we find that the initial evaporation, the induced upflow, and the draining during catastrophic collapse do show steady state behavior, namely that the mass flow rate $\rho A v$ is spatially constant ($\partial_{s} (\rho A v) \approx 0$).  We now show that this assumption holds well for all cases during catastrophic collapse, and holds approximately during evaporation and the induced upflow, particularly for loops with larger expansion factors.

\begin{figure*}
    \begin{minipage}[b]{0.88\linewidth}
    \centering
    \includegraphics[width=0.31\textwidth]{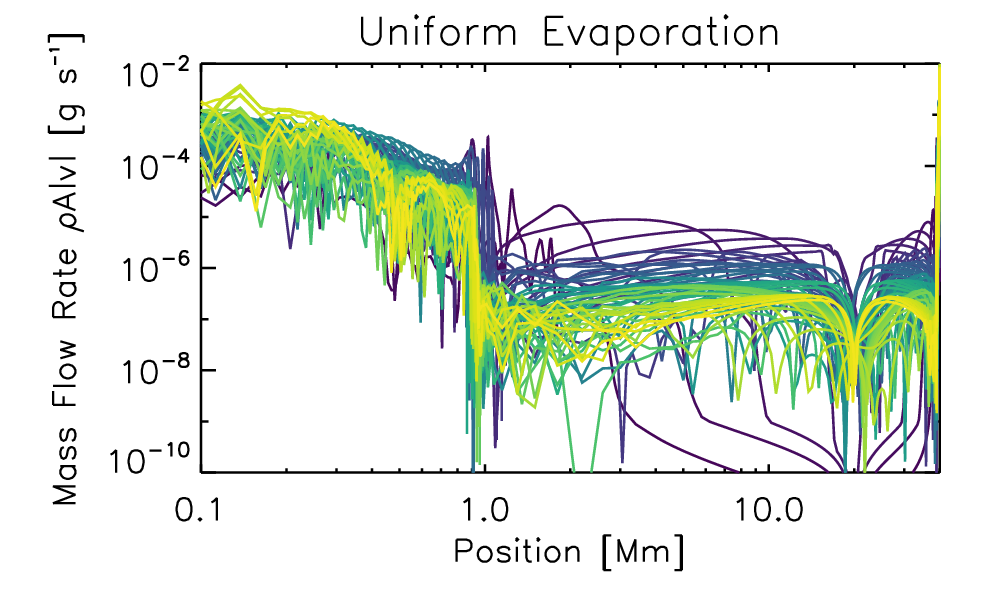}
    \includegraphics[width=0.31\textwidth]{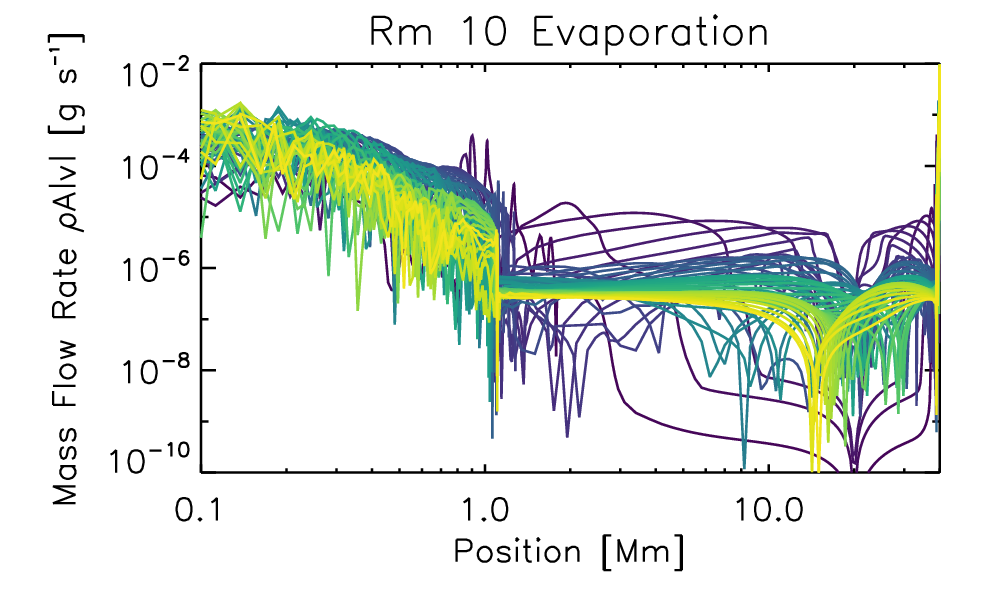}
    \includegraphics[width=0.31\textwidth]{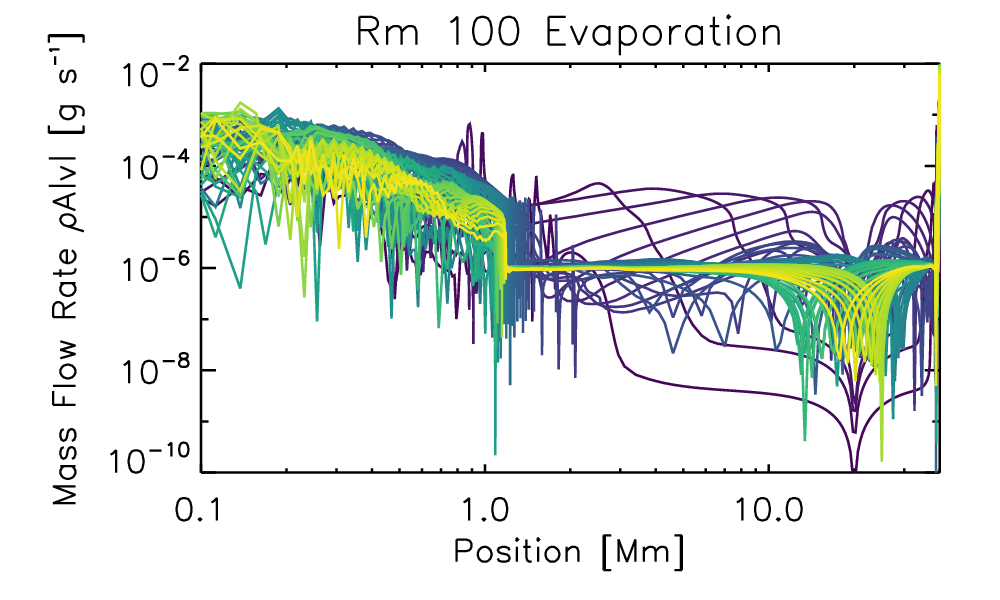}
    \includegraphics[width=0.31\textwidth]{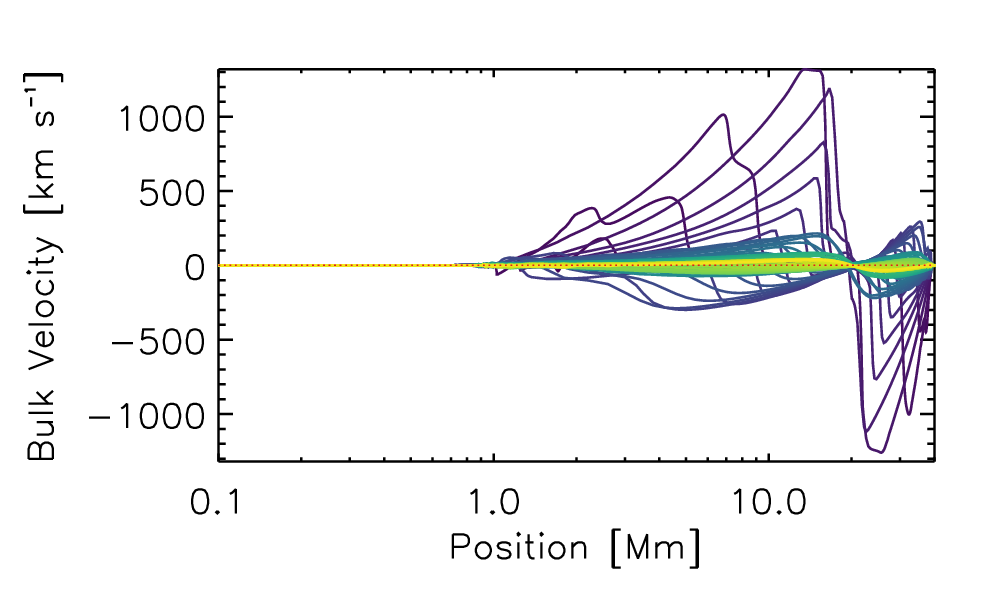}
    \includegraphics[width=0.31\textwidth]{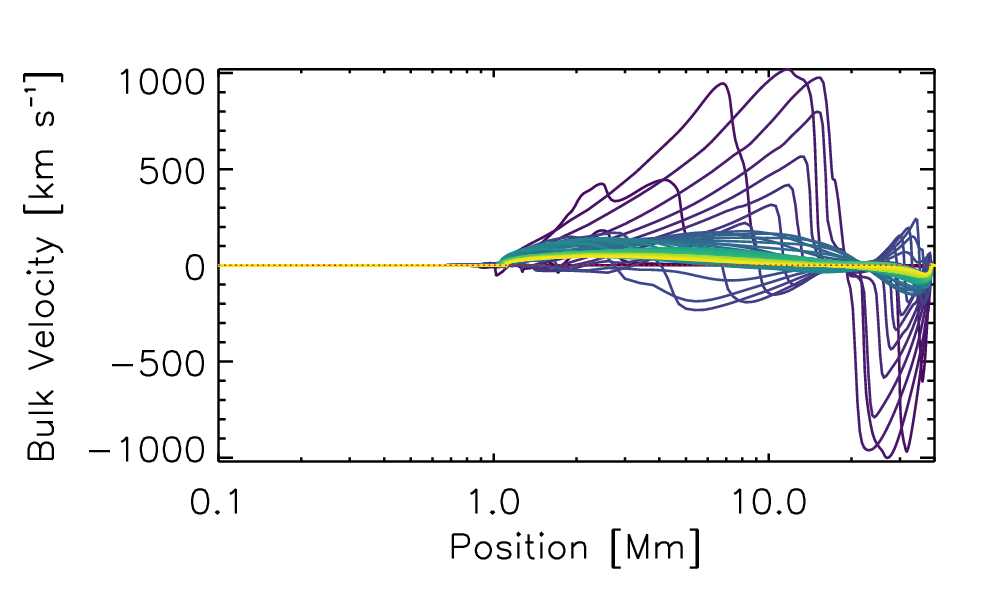}
    \includegraphics[width=0.31\textwidth]{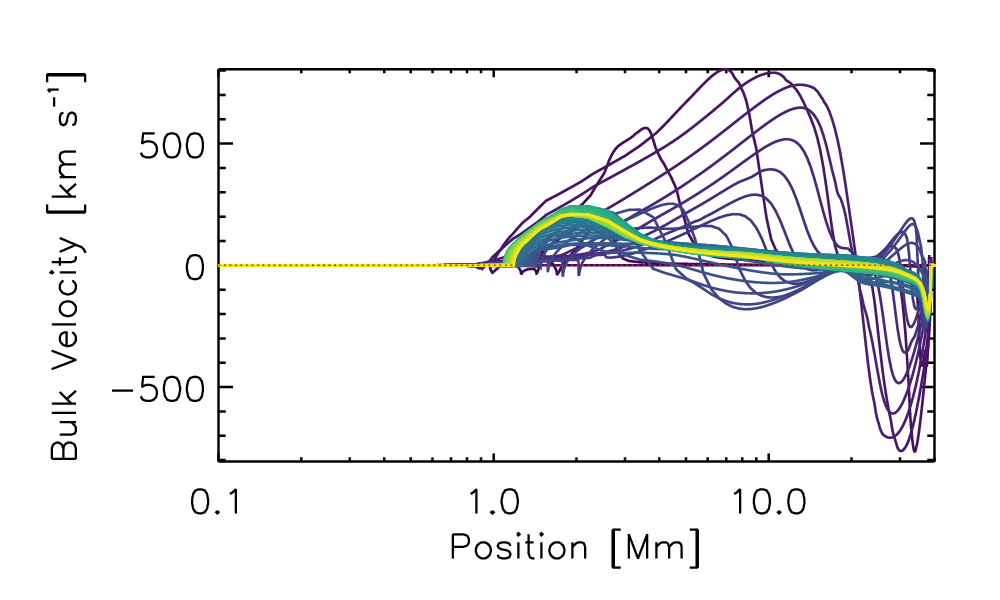}
    \includegraphics[width=0.31\textwidth]{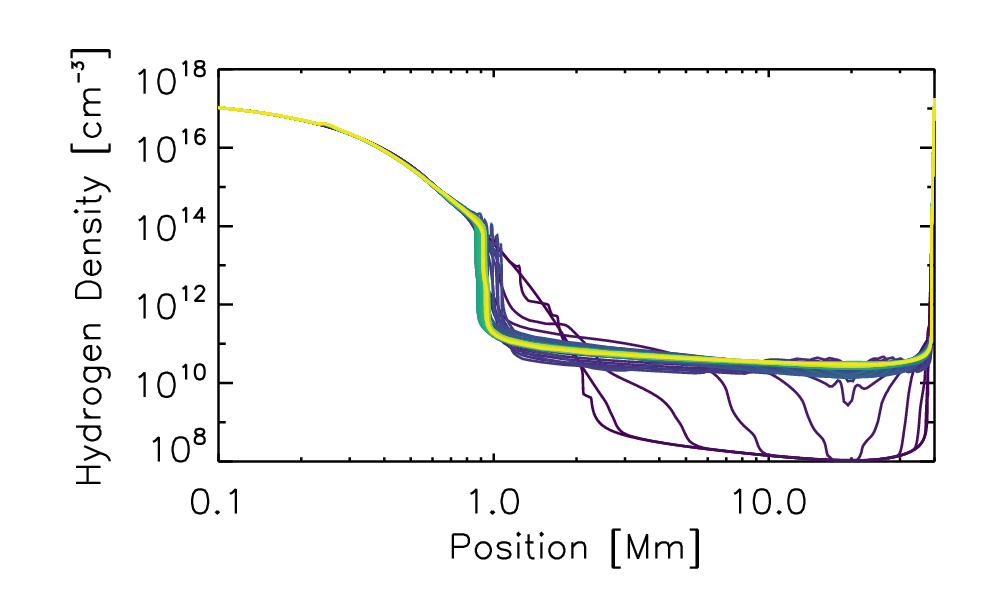}
    \includegraphics[width=0.31\textwidth]{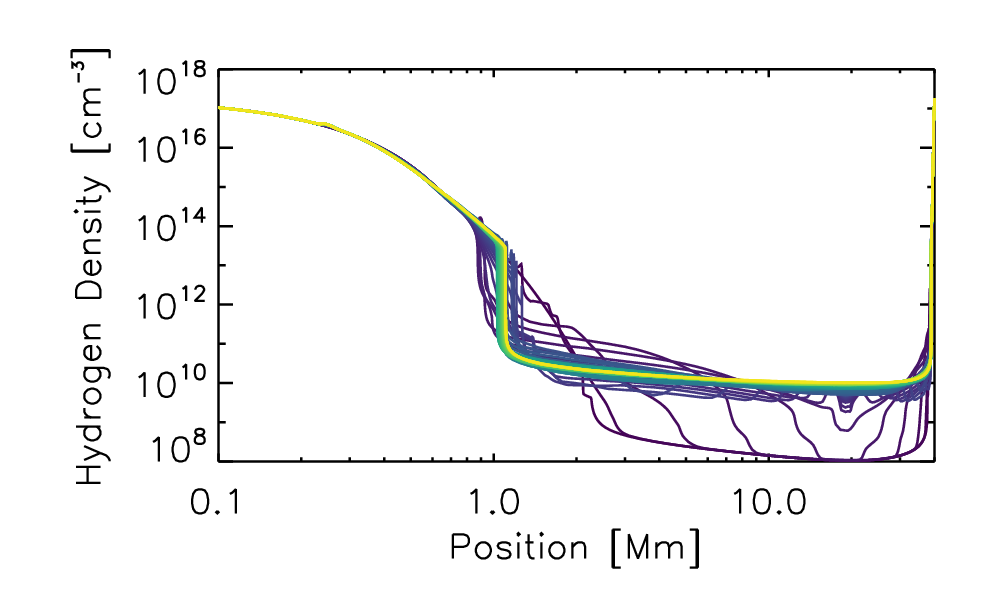}
    \includegraphics[width=0.31\textwidth]{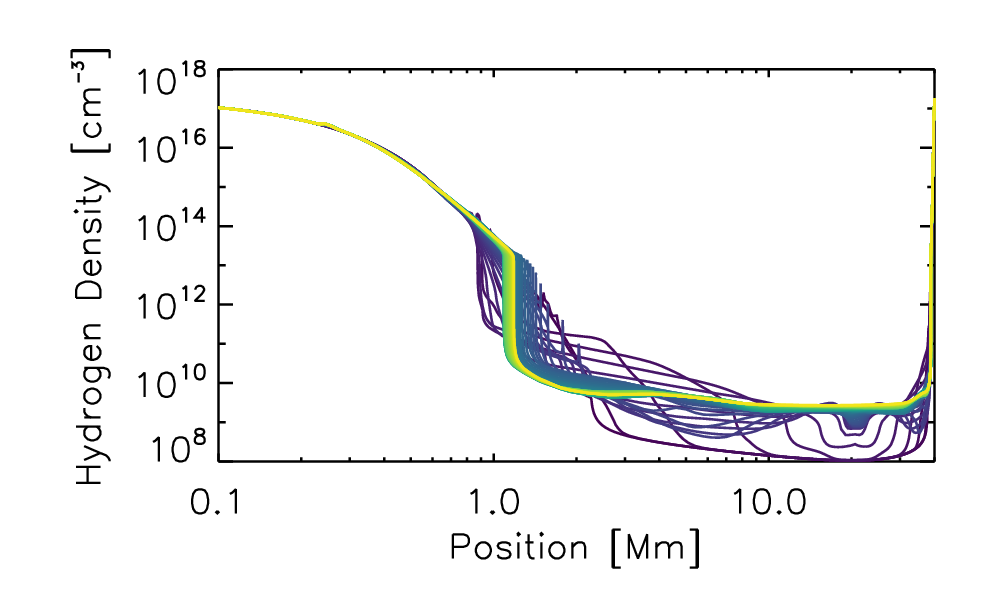}
    \end{minipage}
    \hfill
    \begin{minipage}[b]{0.1\linewidth}
     \includegraphics[width=\textwidth]{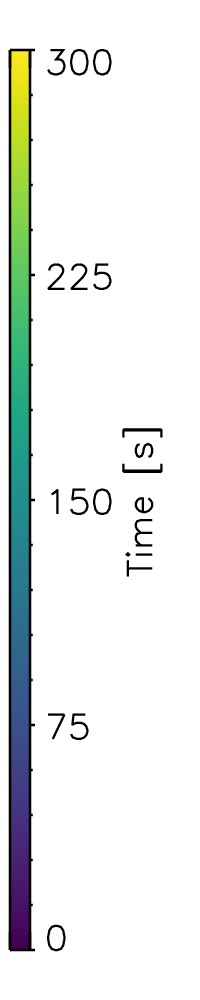}   
    \end{minipage}
    \caption{Flow behavior for the simulations in Figure \ref{fig:heated} during the initial 300 s after the onset of heating.  The top row shows the mass flow rate ($\rho A |v|$) as a function of position, with time at a 5-s cadence from dark blue through yellow.  The middle row shows the bulk velocity $v$ at the same times, with positive flows defined as moving to the right.  The bottom row shows the hydrogen density $n_{H} (\propto \rho)$.  The columns show a uniform area loop, along with expansion factors of 10 and 100 from footpoint to apex, respectively.  With expansion, the evaporative upflows tend towards steady state between the TR and at a point high in the corona ($\partial_{s}(\rho A v) = 0$), and are steadier for larger expansion factors.  The induced upflow seen in the cases with expansion is steady. 
 \label{fig:steady_evap}}
\end{figure*}

In Figure \ref{fig:steady_evap}, we show plots of the mass flow rate, $\rho A |v|$, for the 300 s after the onset of heating (which lasts 20 s) in three of the loops shown in Figure \ref{fig:heated}, along with plots of the bulk velocity and hydrogen density ($= \frac{\rho}{\mu}$) at the same times.  The columns show loops with uniform area (left), and expansions of $R_{m} = 10$ (center) and $R_{m} = 100$ (right) from footpoint to apex.  The colors, ranging from dark blue to yellow in time, show the evolution at a 5-s cadence.  In the cases with expansion, the evaporation initially develops with non-steady flows, but gradually tends towards steady state after 30 s or so, so that the mass flow rate becomes approximately constant from the top of the TR through a point high up in the corona.  With larger expansion, the upflows are steadier.  In contrast, the uniform case is generally not steady at any time, and the flows fade quickly after heating subsides.  Additionally, the induced long-lasting flow that develops in the cases with expansion forms and remains steady for much longer than the heating duration.  The loop is then divided into three sections: the chromosphere, which is never in steady state; the steady-state region, which forms between the TR and some critical point in the corona; and a region around the apex, where the flow ends, and is never in steady state.  Gradually, the steady flow dissipates as the radiation grows stronger, and as a result breaks the parity in Equation \ref{eqn:parity}.

\begin{figure*}
    \begin{minipage}[b]{0.88\linewidth}
    \centering
    \includegraphics[width=0.31\textwidth]{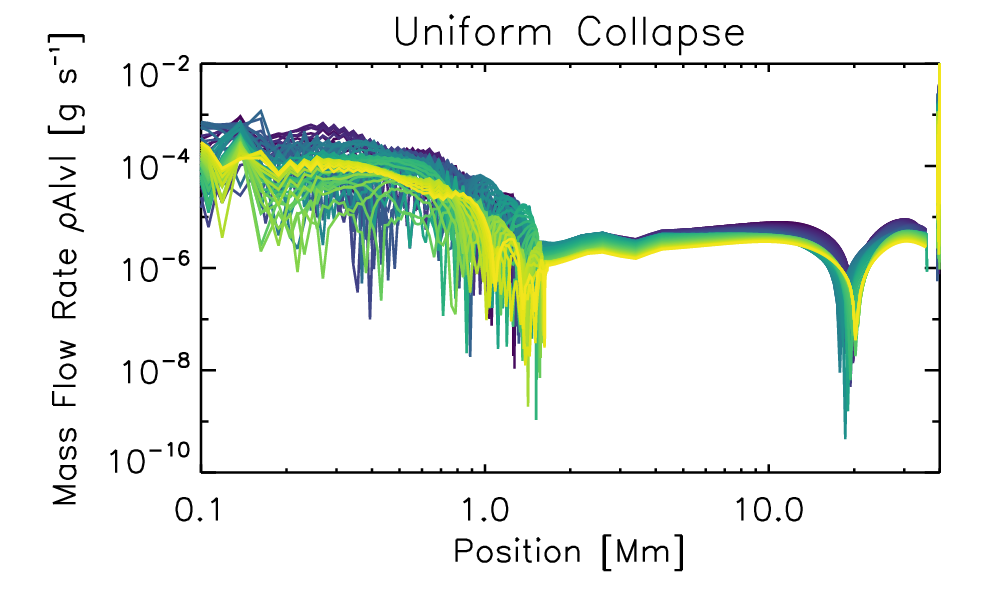}
    \includegraphics[width=0.31\textwidth]{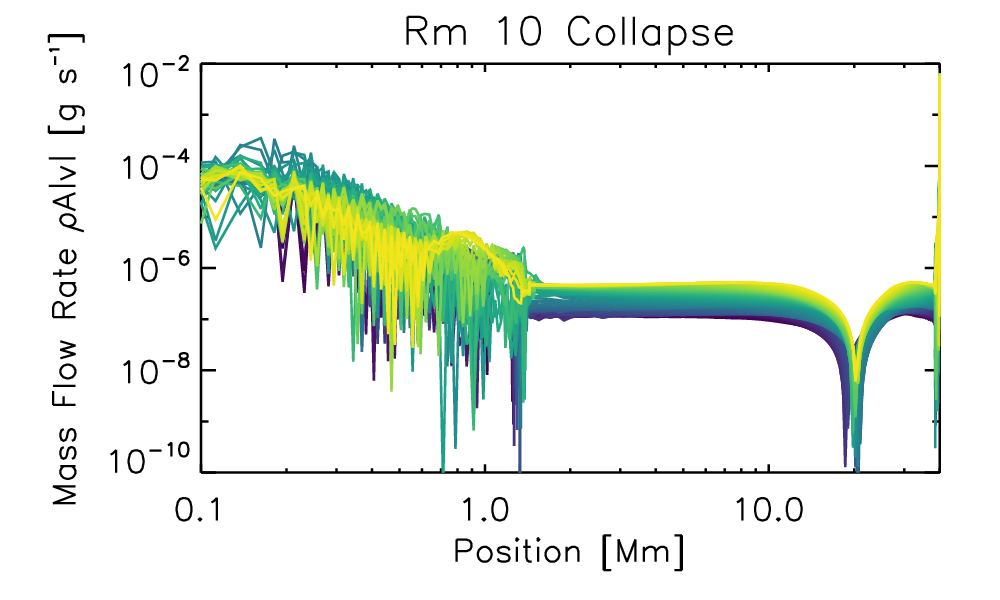}
    \includegraphics[width=0.31\textwidth]{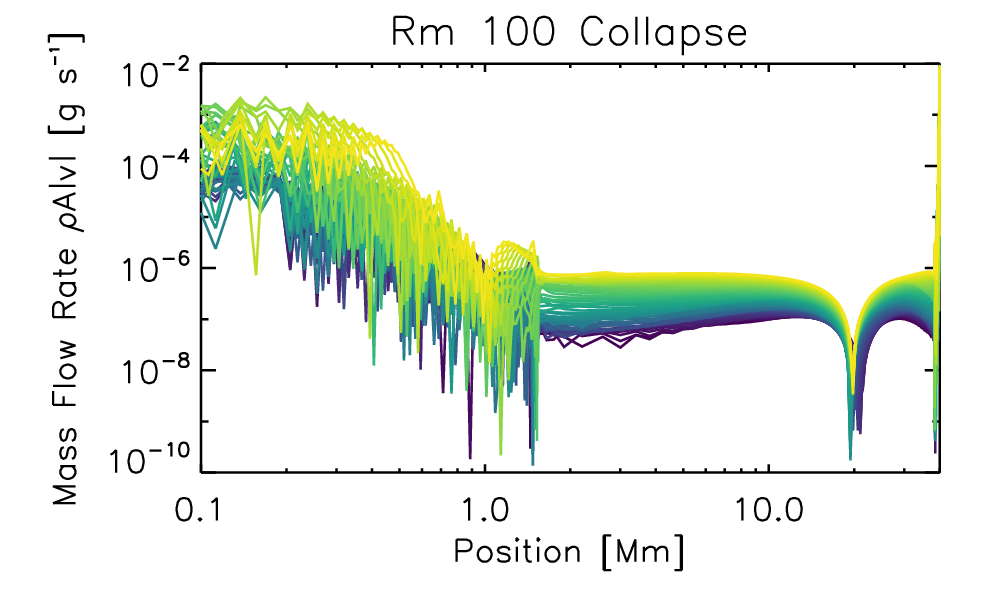}
    \includegraphics[width=0.31\textwidth]{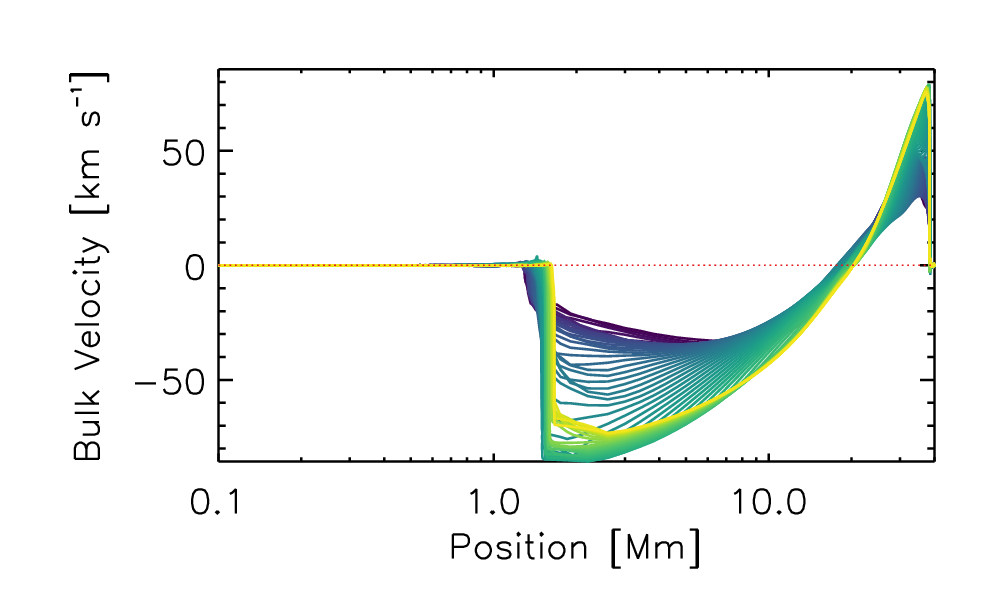}
    \includegraphics[width=0.31\textwidth]{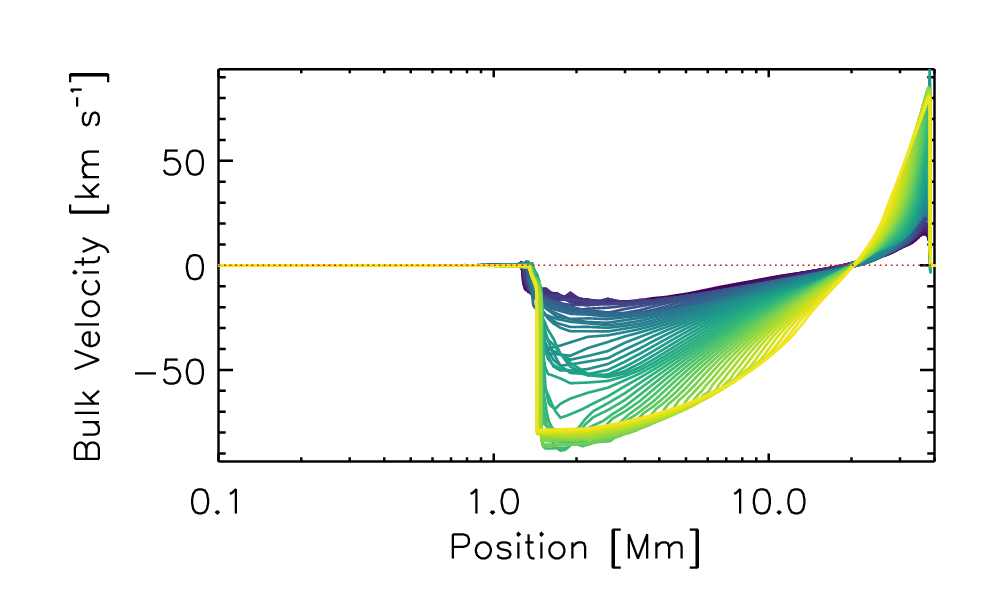}
    \includegraphics[width=0.31\textwidth]{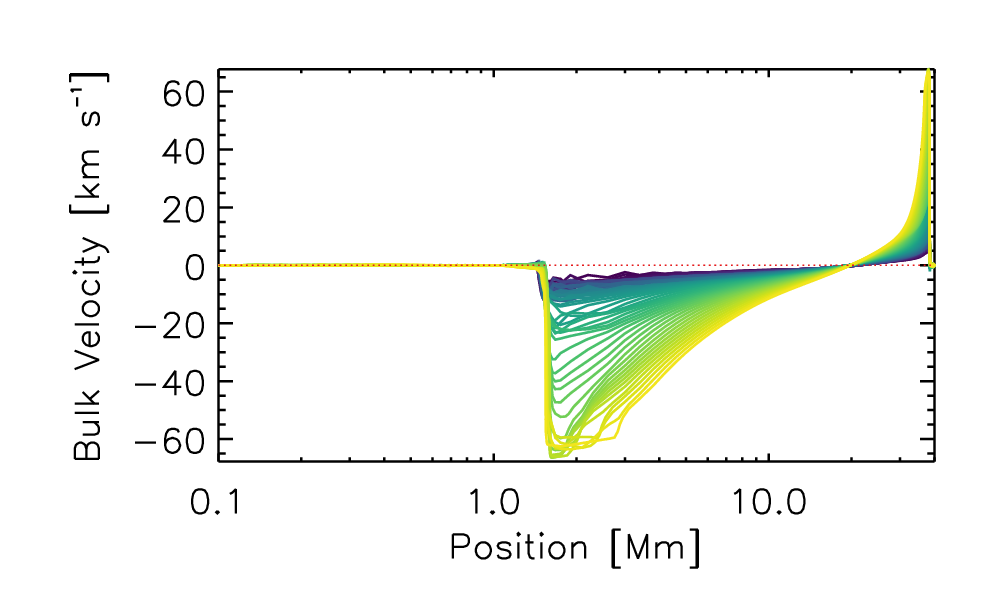}
    \includegraphics[width=0.31\textwidth]{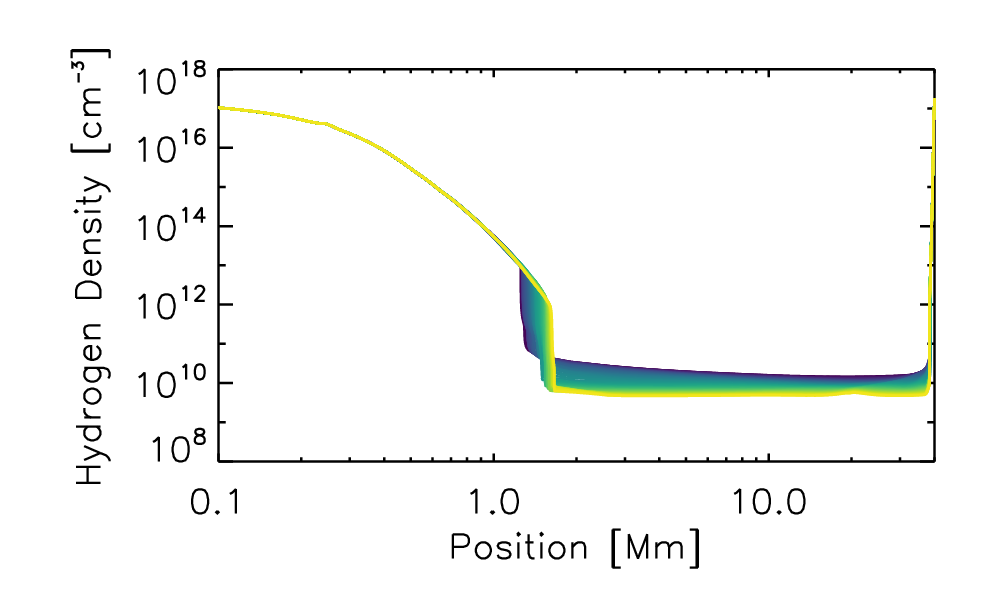}
    \includegraphics[width=0.31\textwidth]{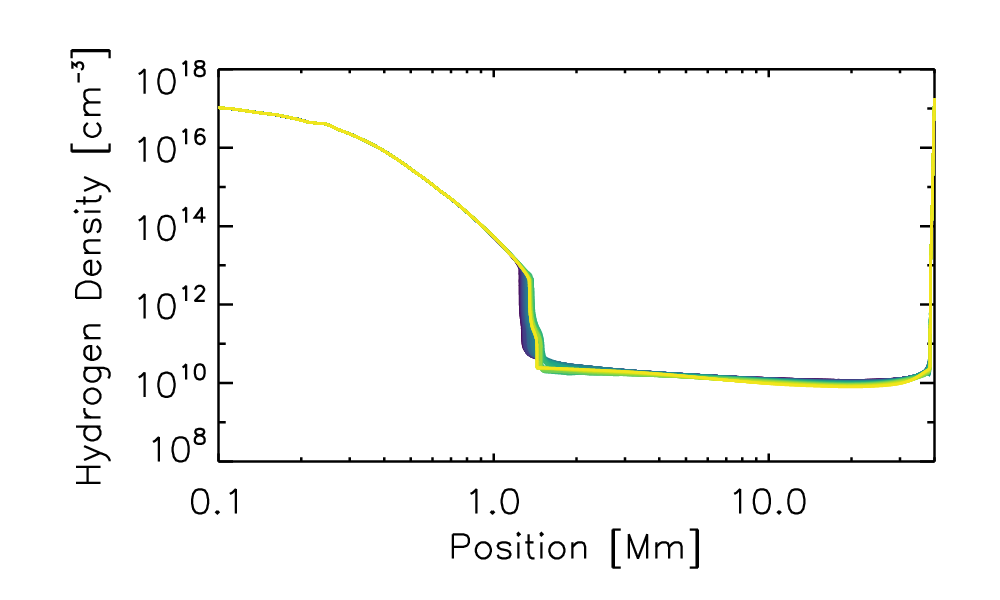}
    \includegraphics[width=0.31\textwidth]{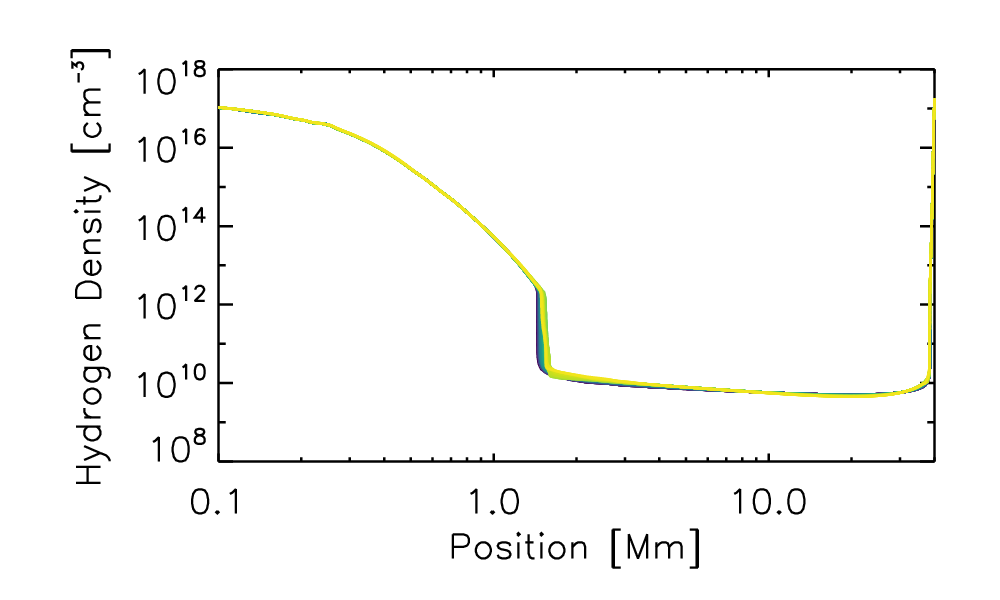}
    \end{minipage}
    \hfill
    \begin{minipage}[b]{0.1\linewidth}
     \includegraphics[width=\textwidth]{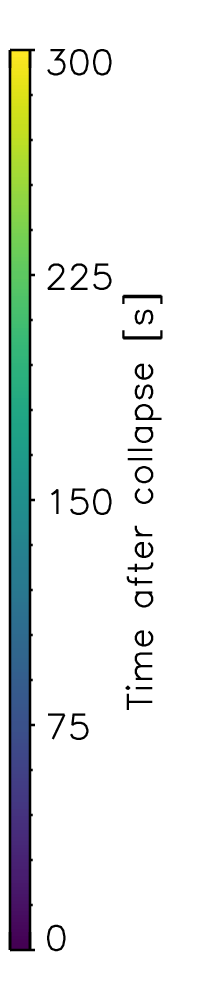}   
    \end{minipage}
    \caption{Flow behavior during the 300 s after the onset of catastrophic collapse ($\approx 900, 1600, 2400$ s into the simulations, respectively), for the same loops as in Figure \ref{fig:heated}.  The catastrophic draining occurs in steady state between the TR and a point near, but not at, the loop apex.  
 \label{fig:steady_coll}}
\end{figure*}

Figure \ref{fig:steady_coll} similarly shows the flow behavior during 300 s near the onset of the catastrophic collapse of each of the loops in Figure \ref{fig:steady_evap} (beginning at 1450, 2000, and 3000 s past heating onset, respectively).  In all three cases, the catastrophic collapse occurs in steady state between the TR through a point near the apex of the loop, with larger expansions having steadier flows.  As before, the flows in the chromosphere are never steady state, while there is also a region near the loop apex where the steady state assumption breaks down.  

\begin{figure*}
    \begin{minipage}[b]{0.88\linewidth}
    \centering
    \includegraphics[width=0.31\textwidth]{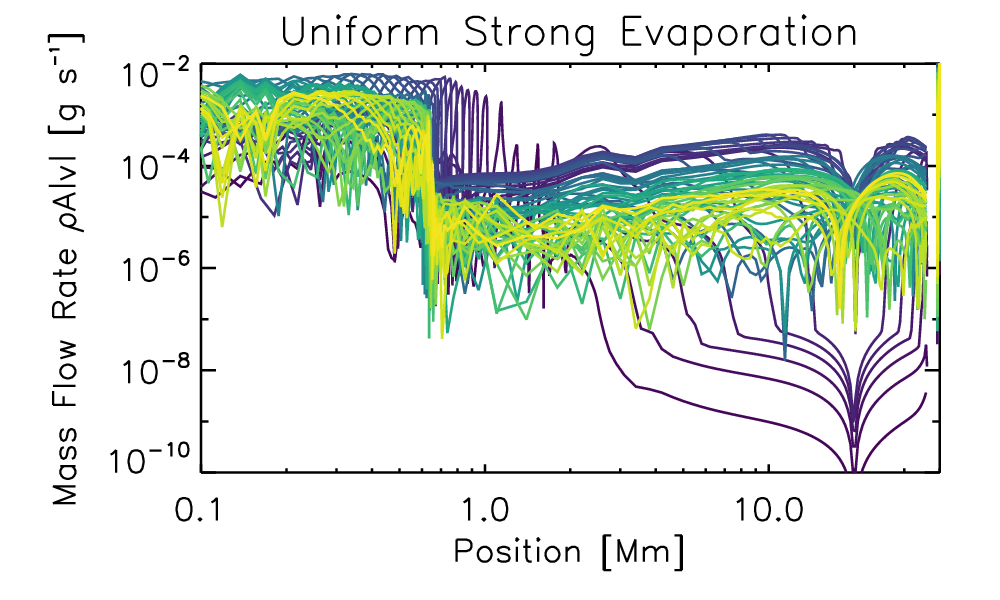}
    \includegraphics[width=0.31\textwidth]{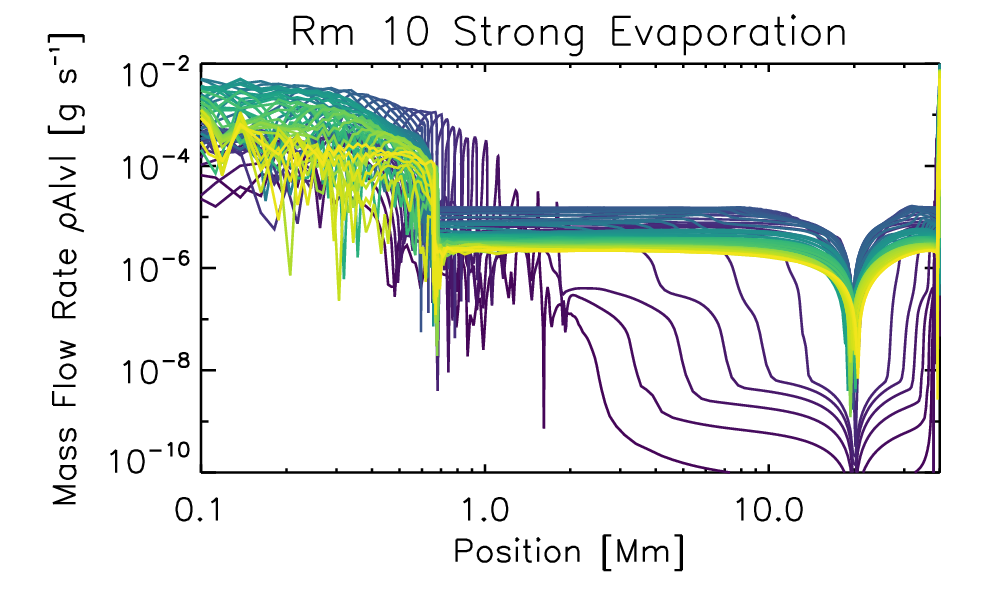}
    \includegraphics[width=0.31\textwidth]{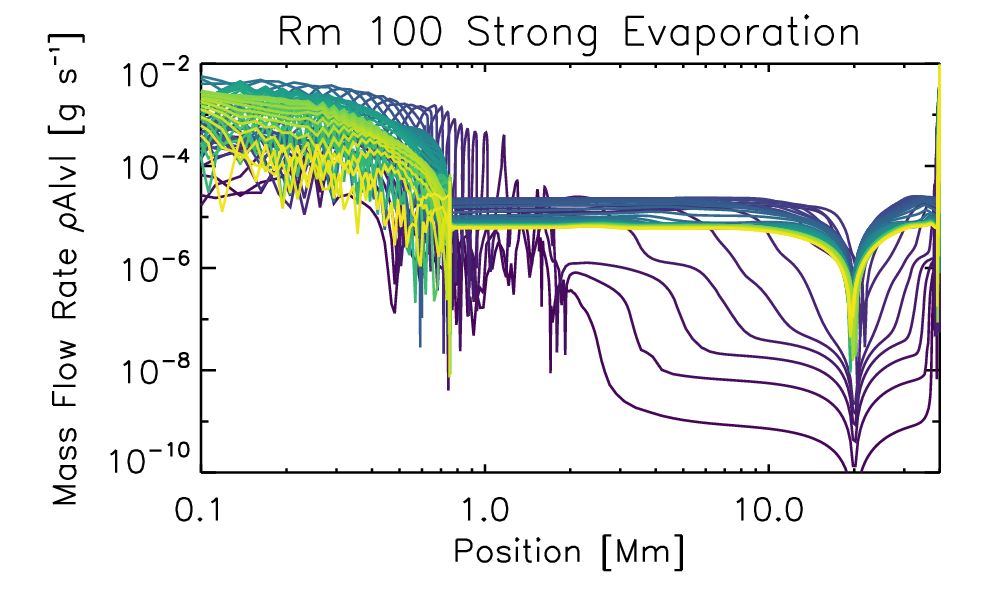}
    \includegraphics[width=0.31\textwidth]{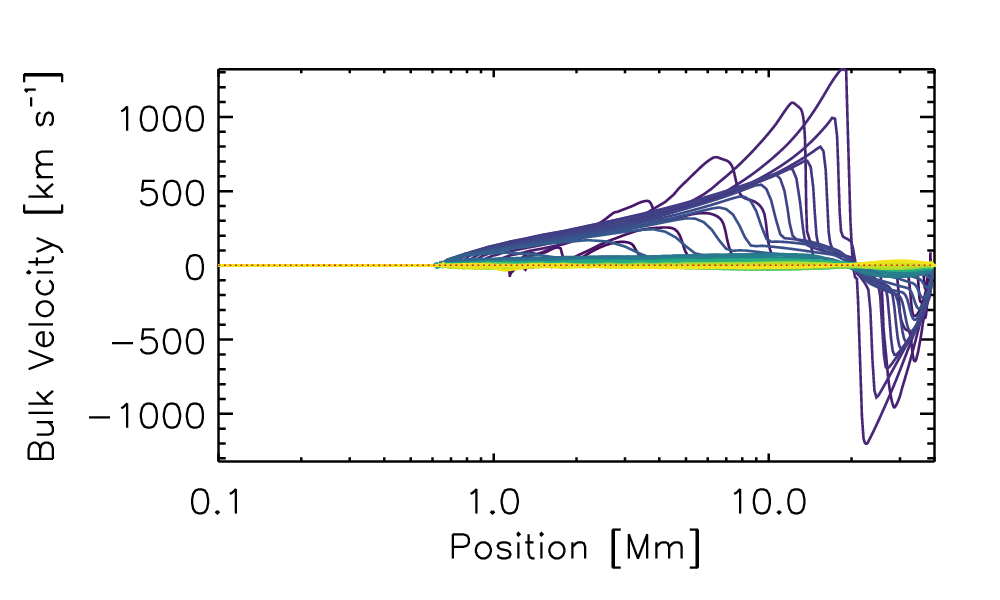}
    \includegraphics[width=0.31\textwidth]{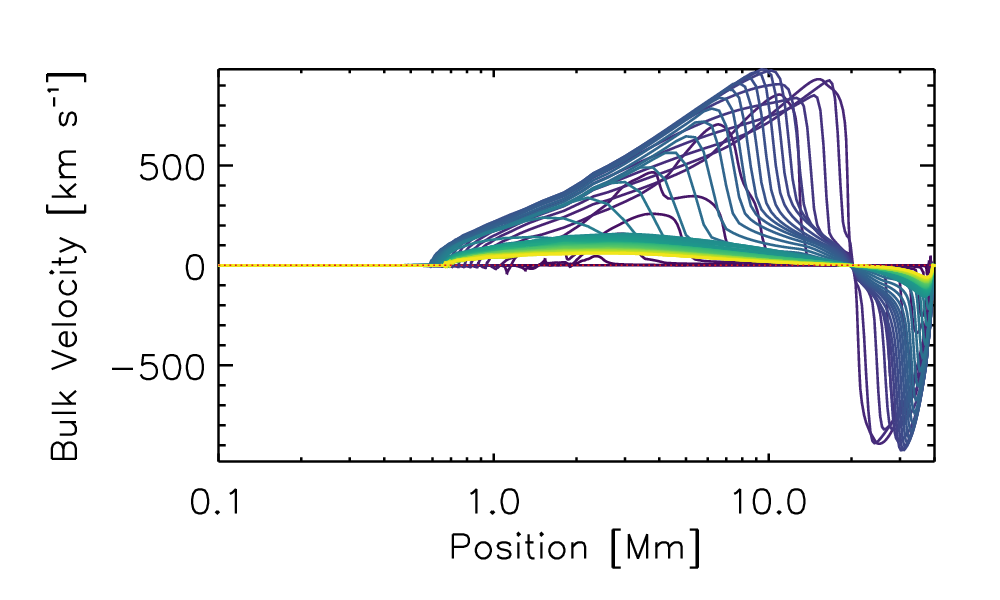}
    \includegraphics[width=0.31\textwidth]{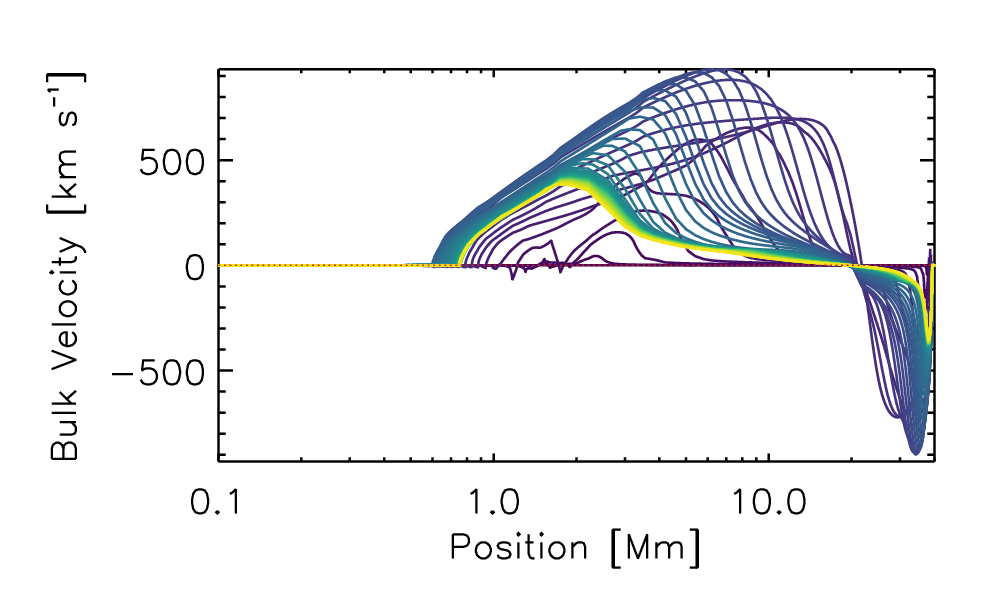}
    \includegraphics[width=0.31\textwidth]{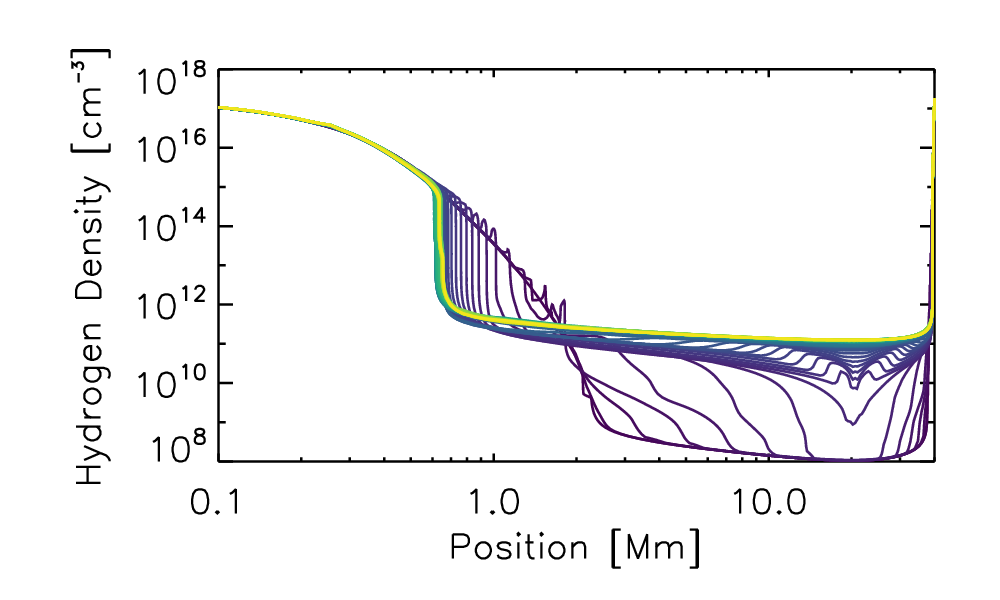}
    \includegraphics[width=0.31\textwidth]{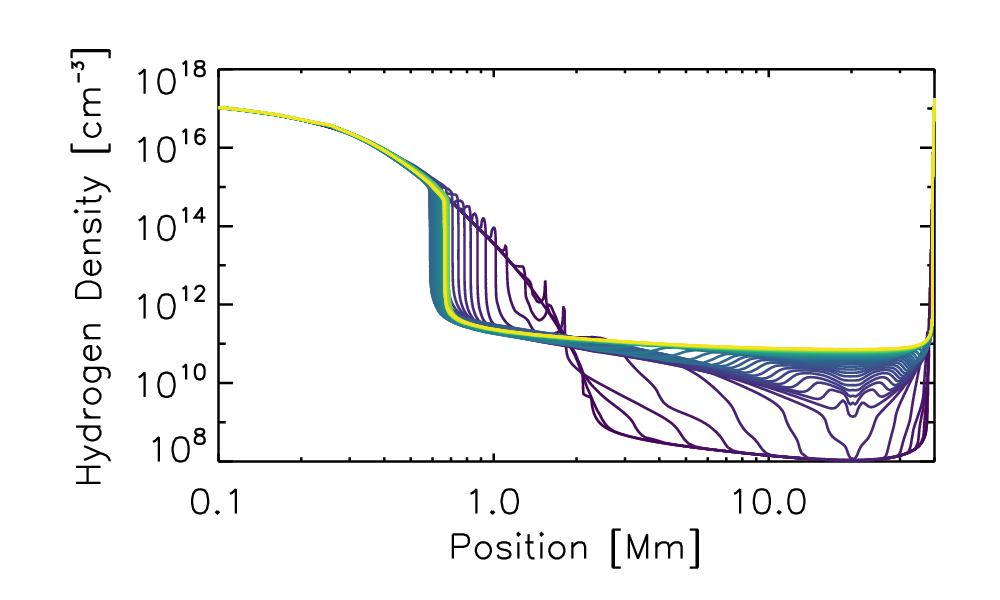}
    \includegraphics[width=0.31\textwidth]{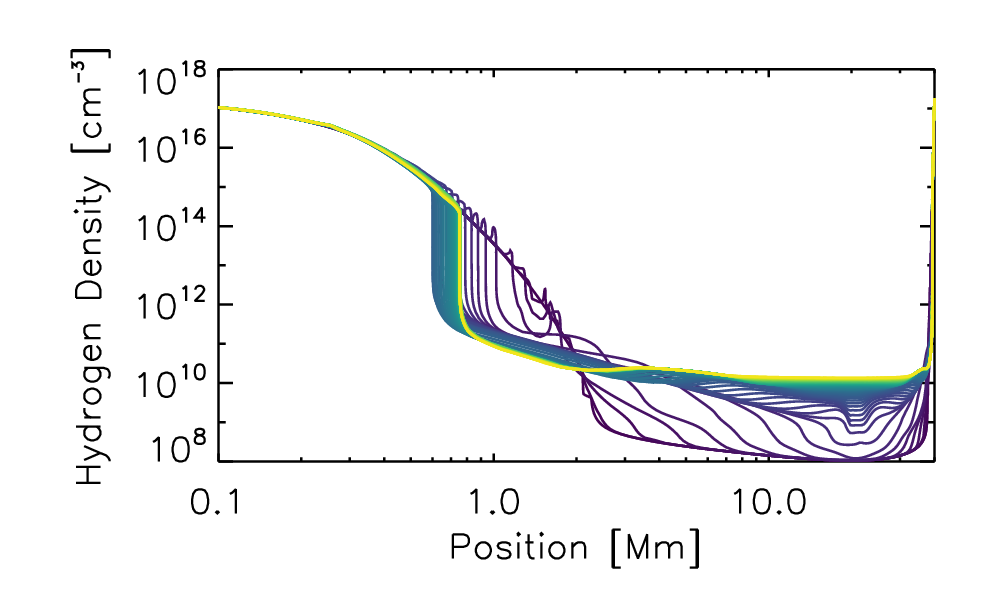}
    \end{minipage}
    \hfill
    \begin{minipage}[b]{0.1\linewidth}
     \includegraphics[width=\textwidth]{evap_colorbar.png}   
    \end{minipage}
    \caption{Flow behavior for the simulations in Figure \ref{fig:strongbeam}, strongly heated for 100 s.  Similar to Figure \ref{fig:steady_evap}.  For the stronger heating events here, and thus stronger evaporation, the flows are significantly steadier.  The induced upflows form once again in the cases with area expansion.  
 \label{fig:strong_evap}}
\end{figure*}

Since the heating was relatively modest in these simulations, we additionally show the flow properties during a stronger evaporation event, for the simulations in Figure \ref{fig:strongbeam}, for 100 s of beam heating.  Figure \ref{fig:strong_evap} shows the flow properties of these simulations in detail.  The mass flows in all three cases quickly tend to a steady state flow, and the cases with expansion once again show a long-lasting induced upflow.  Comparing these simulations to Figure \ref{fig:steady_evap}, we see that the stronger heating produces relatively steadier flows, and the steady state is reached sooner.  

\begin{figure*}
    \centering
    \includegraphics[width=0.31\textwidth]{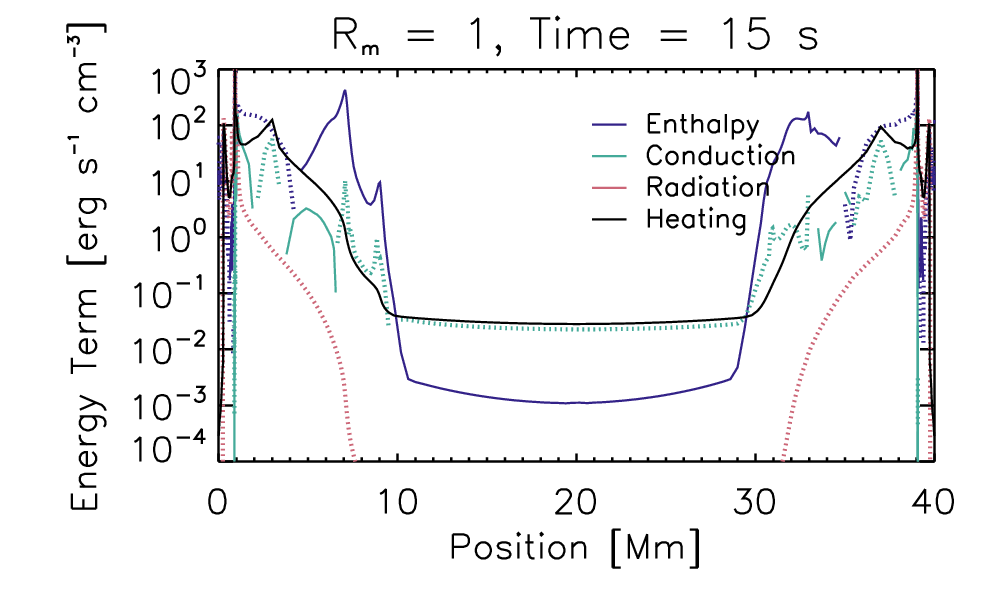}
    \includegraphics[width=0.31\textwidth]{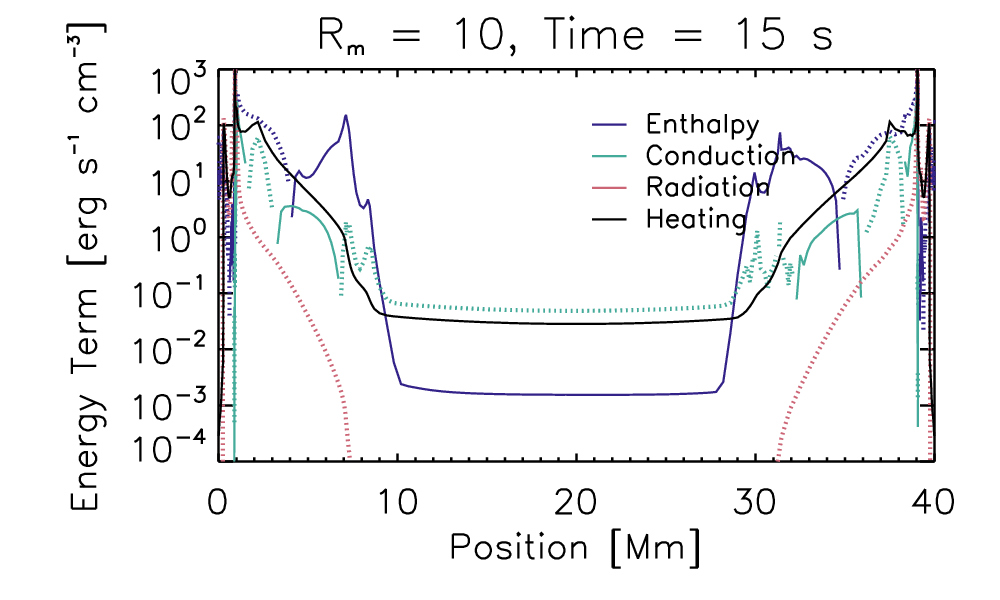}
    \includegraphics[width=0.31\textwidth]{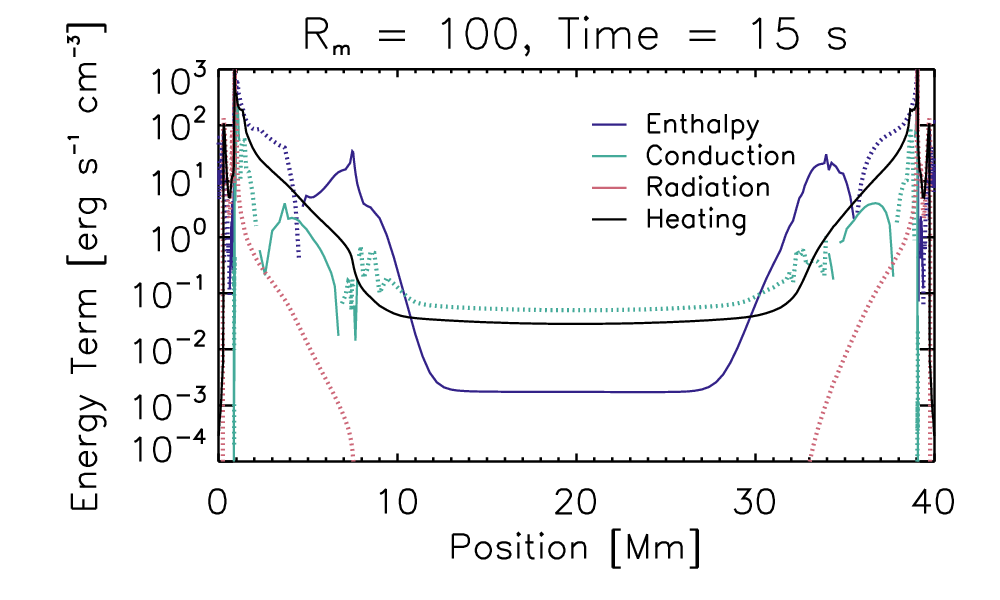}
    \includegraphics[width=0.31\textwidth]{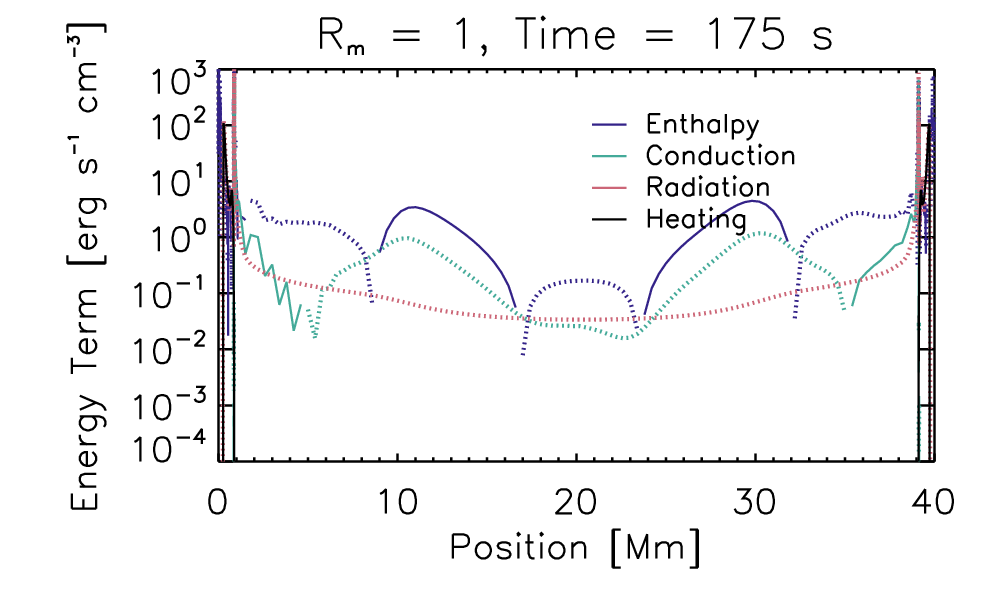}
    \includegraphics[width=0.31\textwidth]{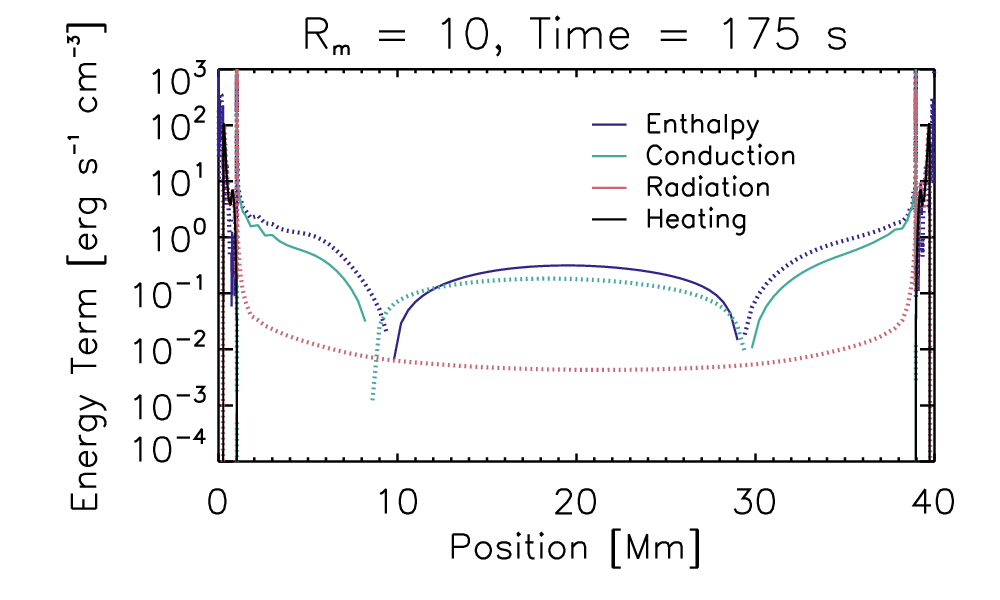}
    \includegraphics[width=0.31\textwidth]{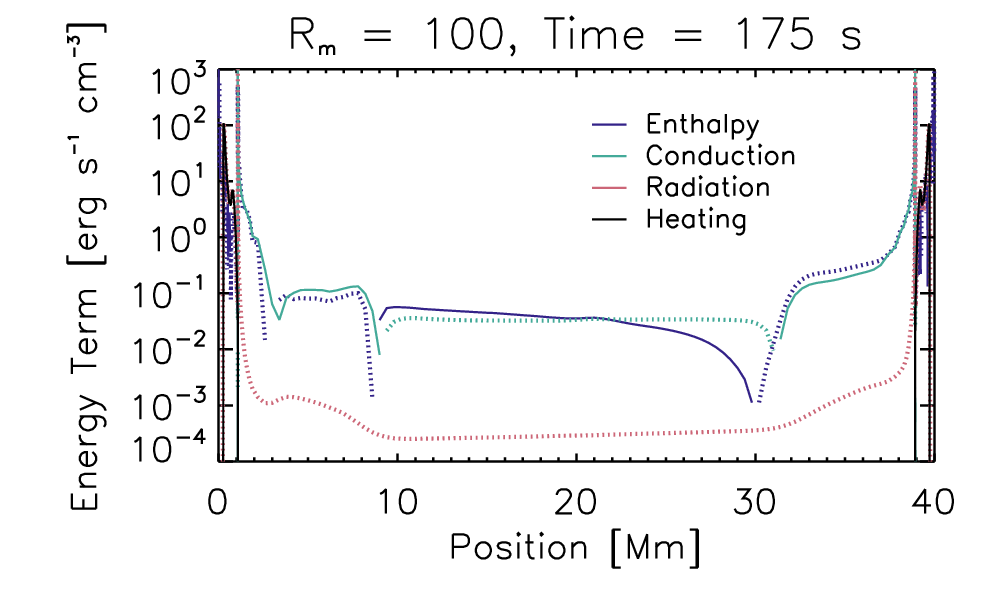}
    \includegraphics[width=0.31\textwidth]{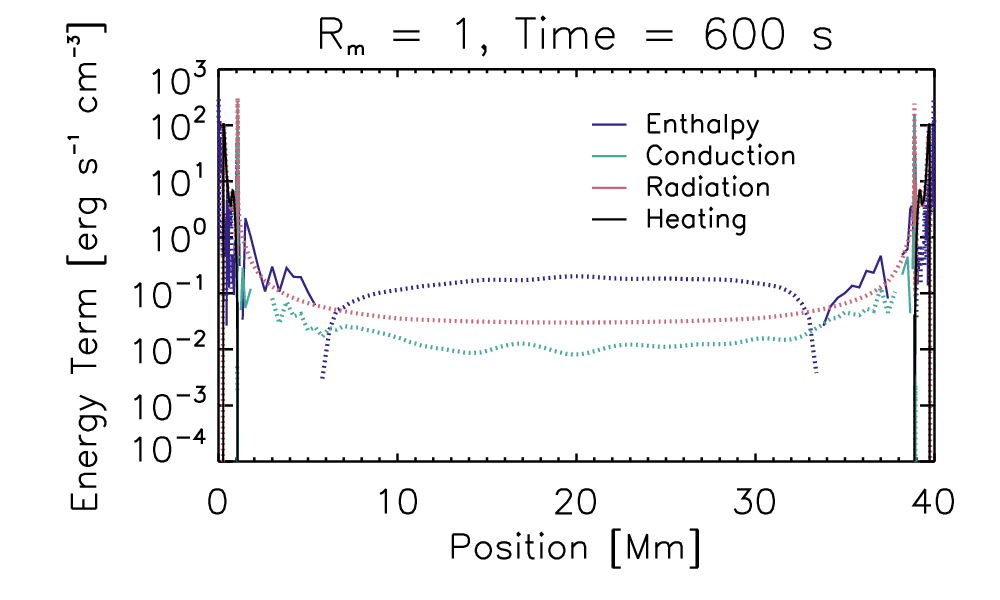}
    \includegraphics[width=0.31\textwidth]{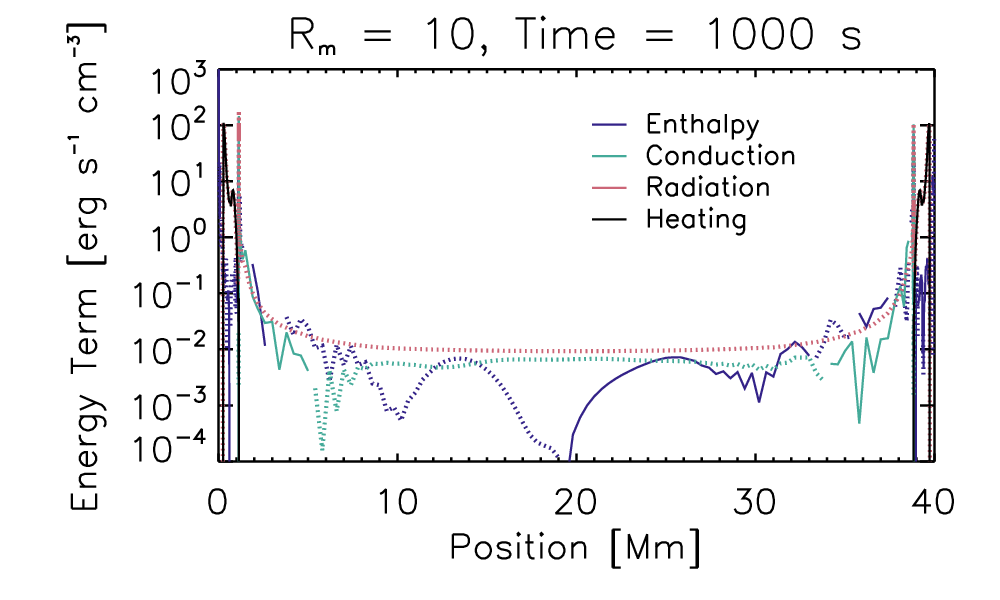}
    \includegraphics[width=0.31\textwidth]{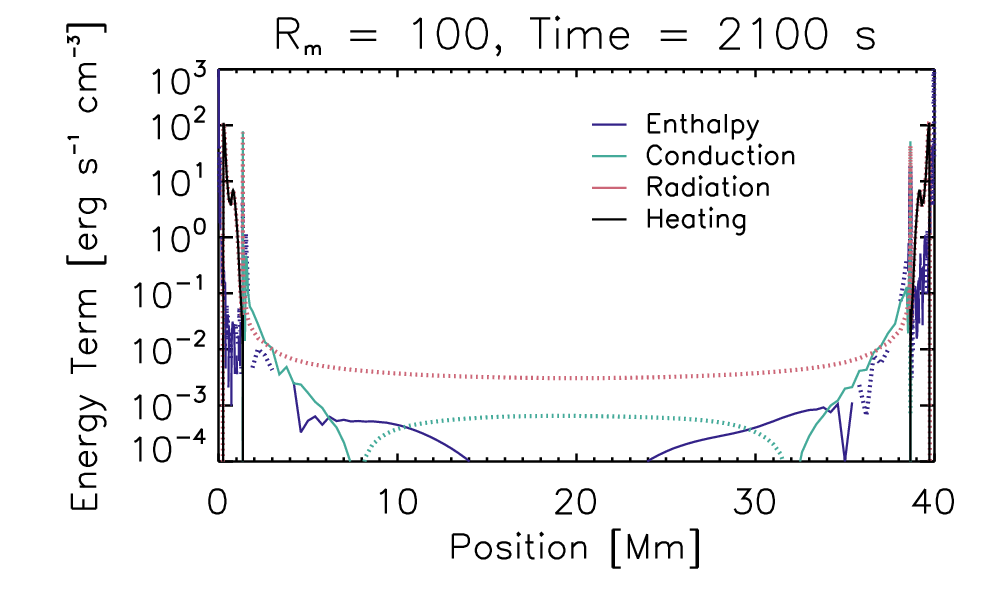}
    \includegraphics[width=0.31\textwidth]{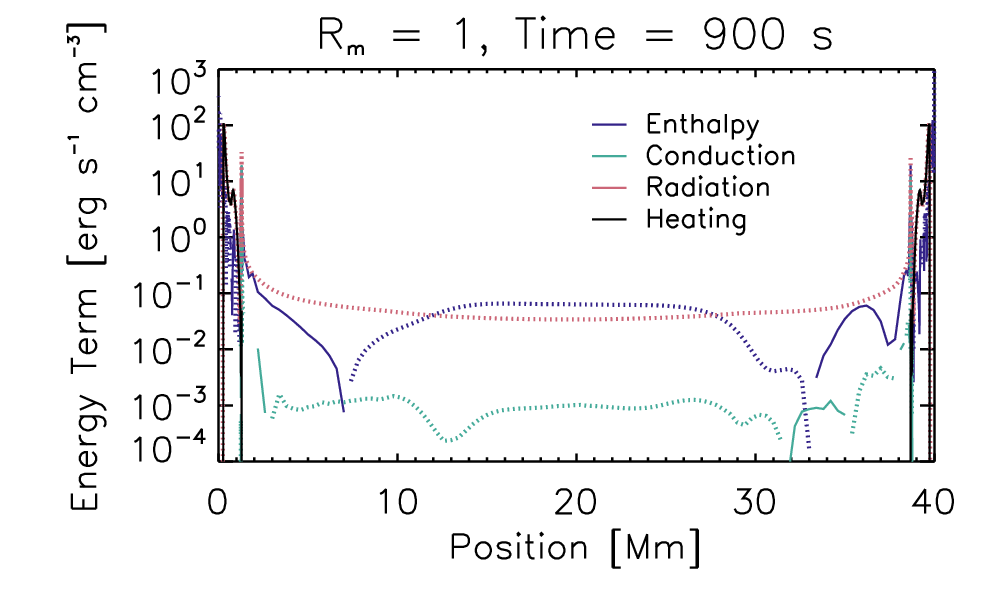}
    \includegraphics[width=0.31\textwidth]{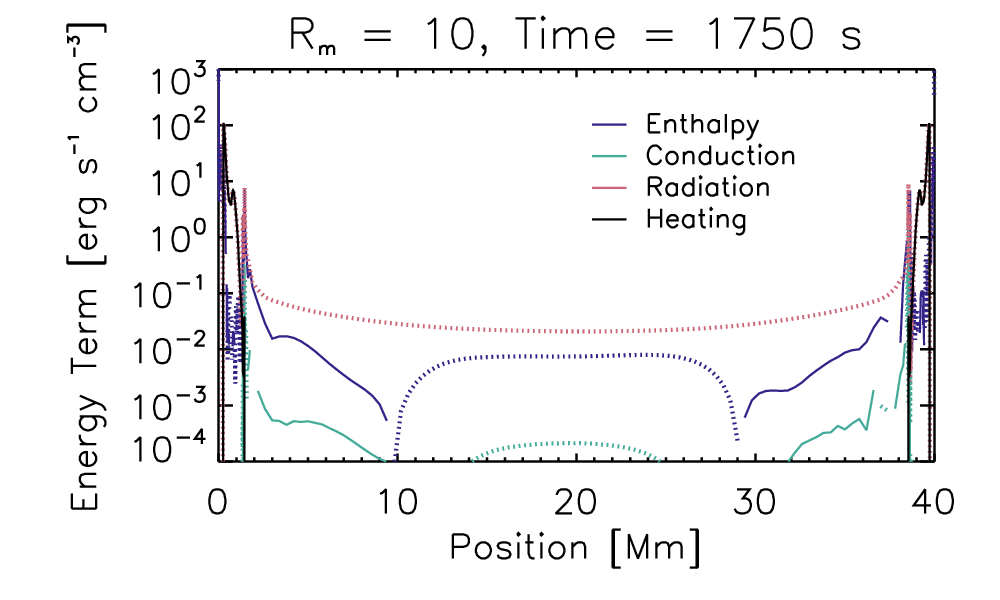}
    \includegraphics[width=0.31\textwidth]{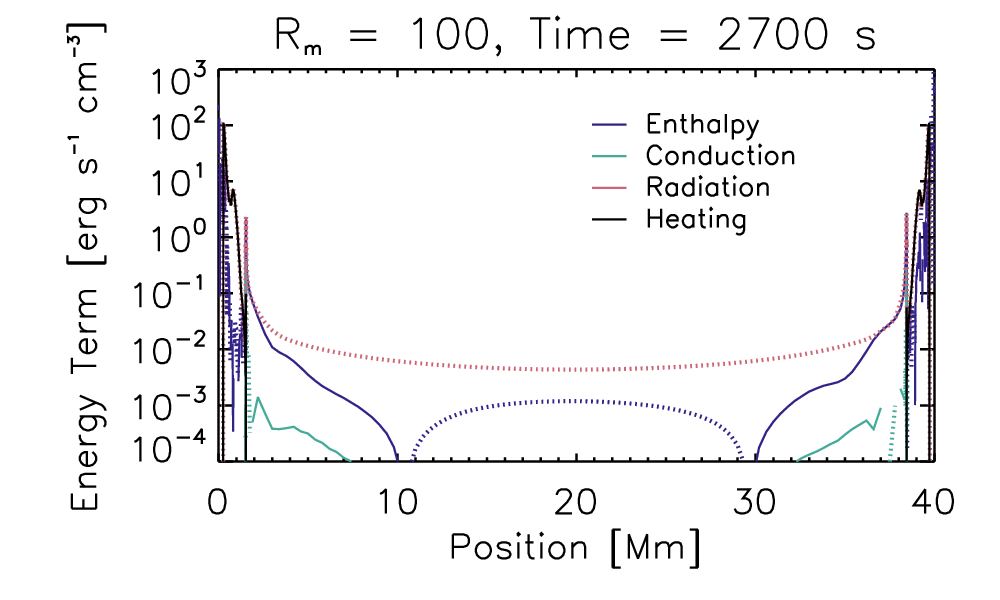}
    \caption{The evolution of the energy terms for enthalpy, thermal conduction, radiative losses (see Equation \ref{eqn:new_balance}), and heating for the three simulations in Figure \ref{fig:steady_evap}.  The columns show the loops with expansion factors $R_{m} =\ $1, 10, 100, respectively.  The rows show four times respectively corresponding approximately to the heating period, early in the cooling phase, the late cooling phase, and after the onset of catastrophic collapse.  Solid lines indicate an energy term is positive, while dotted lines indicate it is negative.
 \label{fig:energy_terms}}
\end{figure*}

In Figure \ref{fig:energy_terms}, we briefly examine the evolution of the energy terms for enthalpy flux (blue), conduction (teal), radiation (pink), and heating (black).  In each case, we add together the contributions to both the electron and hydrogen energy equations in HYDRAD.  The three columns show the three simulations examined in Figures \ref{fig:heated} and \ref{fig:steady_evap}, corresponding respectively to expansion factors $R_{m} =\ $1, 10, 100.  The four rows then show the evolution of the energy terms in each case, at times that correspond to the heating phase, early in the cooling phase, late in the cooling phase, and after the onset of catastrophic collapse.  Solid lines are used to indicate where each term is positive, while dotted lines indicate where a term is negative.

During the heating phase (first row), all three simulations are similar.  The heating terms are positive everywhere, and drive an upflow into the loop.  The enthalpy flux is therefore positive everywhere in the corona, as it fills the loop.  Conduction carries energy out of the corona, while the radiation is completely negligible.  After the heating ceases shortly into the cooling phase (second row), the simulations diverge.  The cases with expansion have a large, positive enthalpy flux centered at the loop apex, approximately balancing the losses due to conduction, while the radiation term is negligible.  This is the sustained upflow, showing that there is indeed an approximate balance between conduction and enthalpy (Equation \ref{eqn:parity}).  In the uniform case, however, the enthalpy is negative at the apex, as the loop has begun to drain.  Late in the cooling phase (third row), the sustained upflows have ceased in the expanding cases, and the loops are now cooling as radiation and conduction slowly carry energy out of the corona.  The uniform loop has continued to drain, with a strong negative enthalpy flux over the whole corona dominating the evolution at this time.  Finally, after the onset of catastrophic collapse (fourth row), all cases show both a strong radiative loss and enthalpy flux carrying energy rapidly out of the corona.

\begin{figure*}
    \centering
    \includegraphics[width=0.31\textwidth]{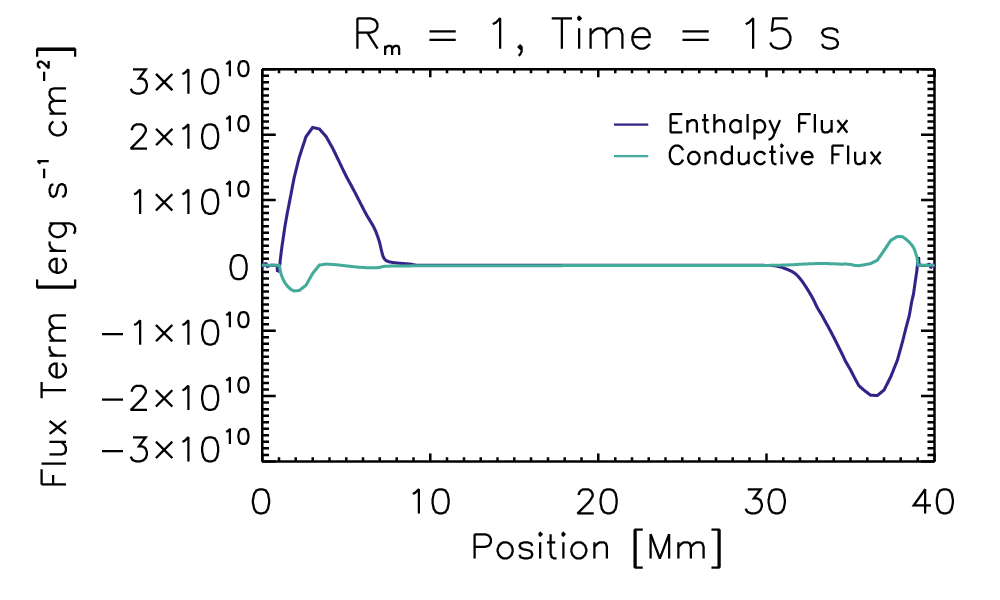}
    \includegraphics[width=0.31\textwidth]{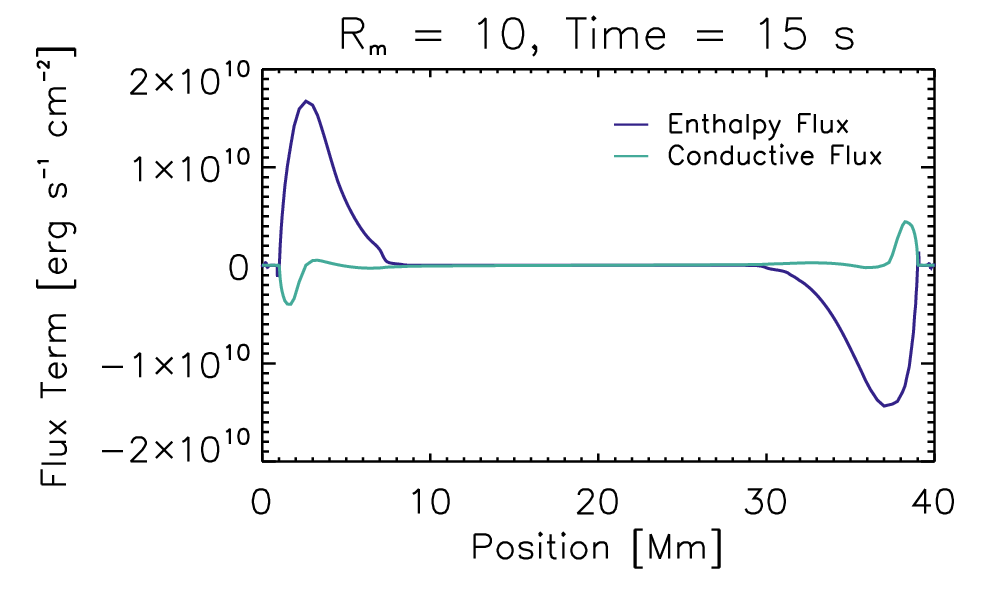}
    \includegraphics[width=0.31\textwidth]{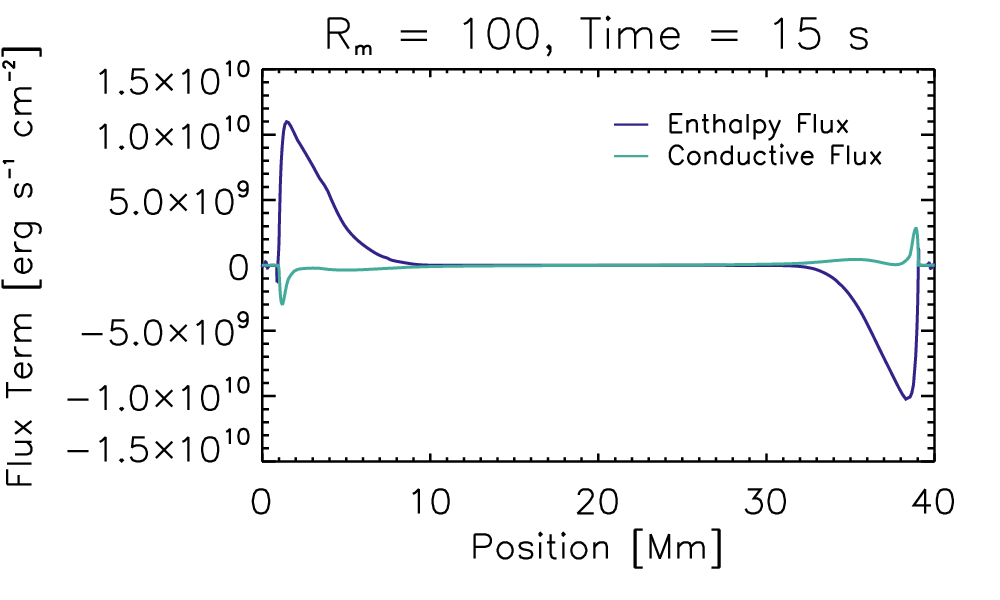}
    \includegraphics[width=0.31\textwidth]{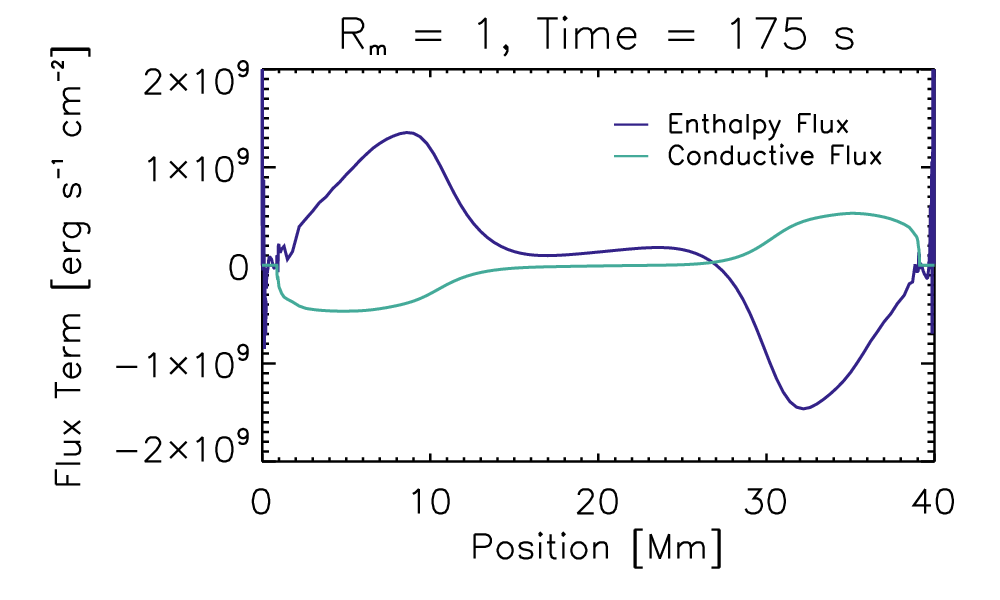}
    \includegraphics[width=0.31\textwidth]{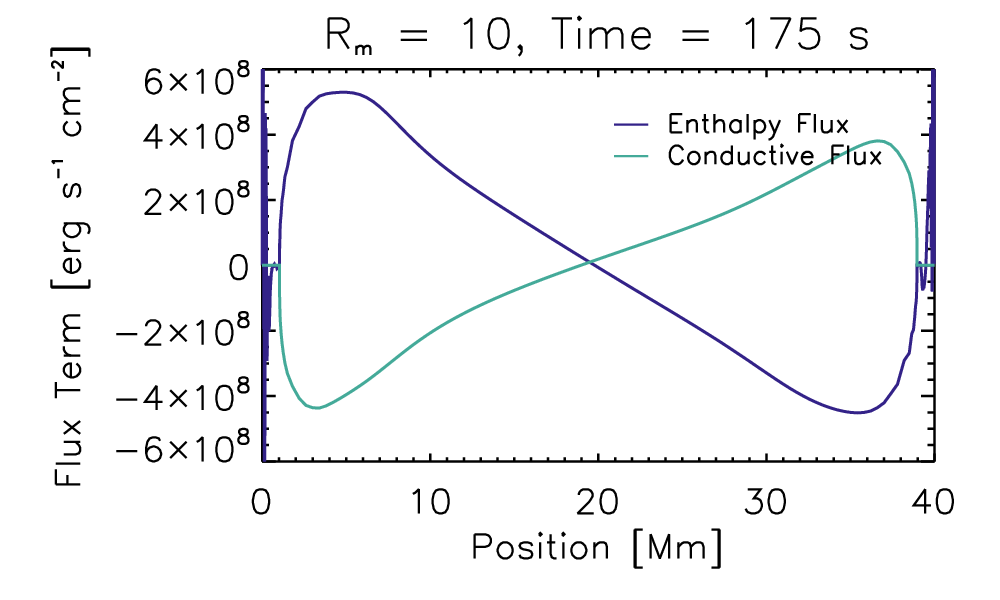}
    \includegraphics[width=0.31\textwidth]{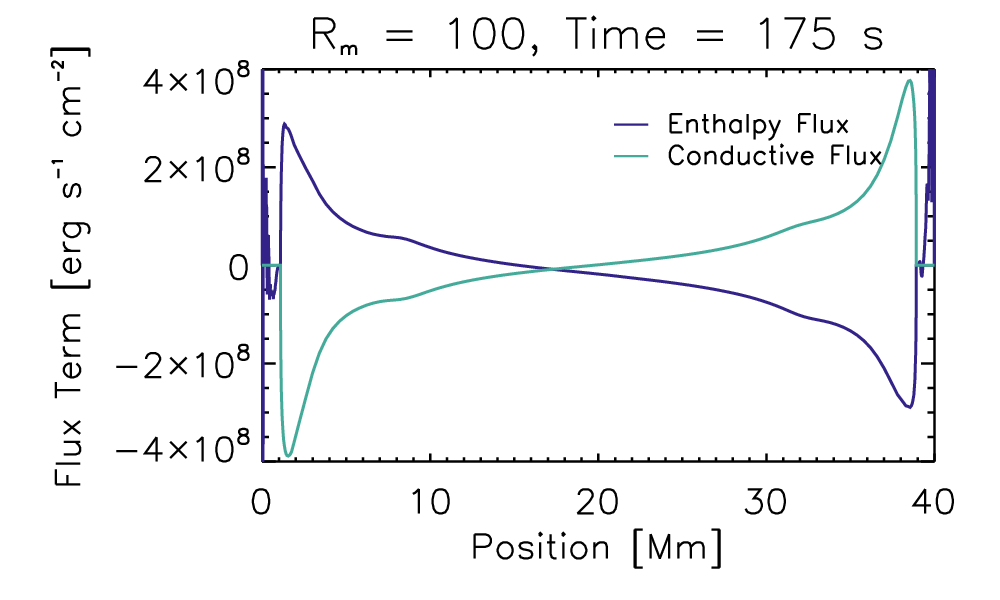}
    \includegraphics[width=0.31\textwidth]{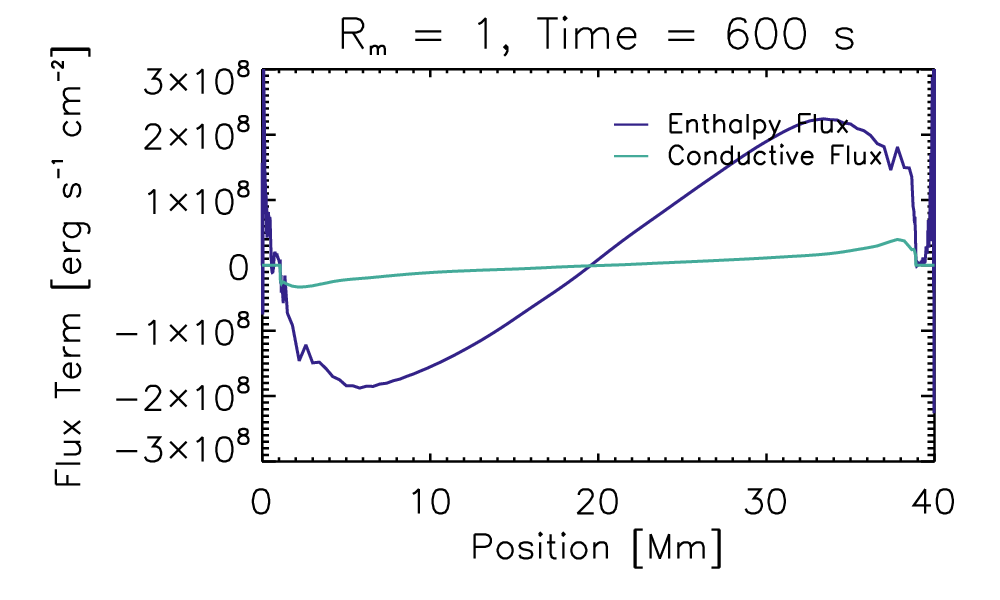}
    \includegraphics[width=0.31\textwidth]{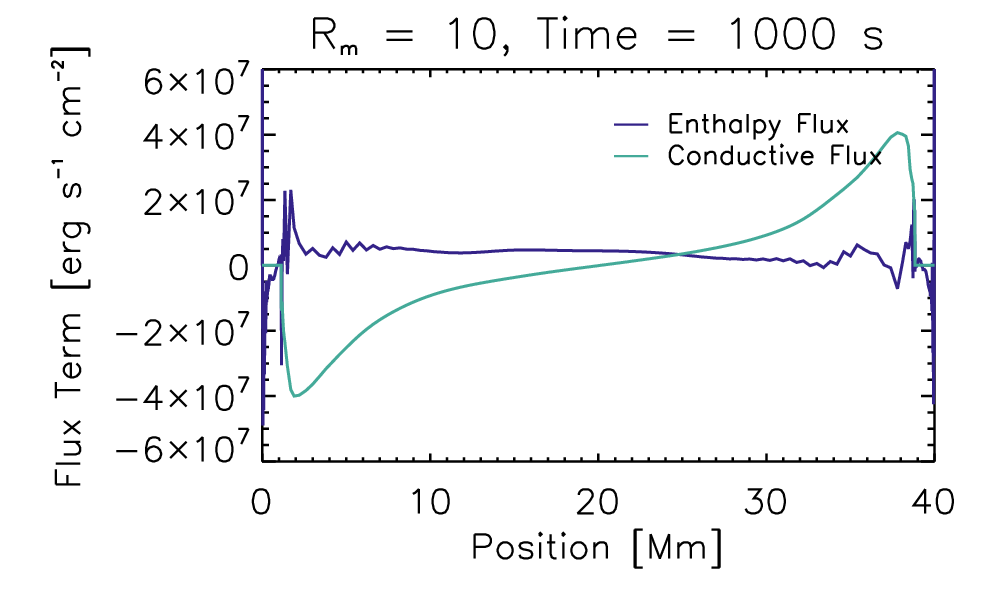}
    \includegraphics[width=0.31\textwidth]{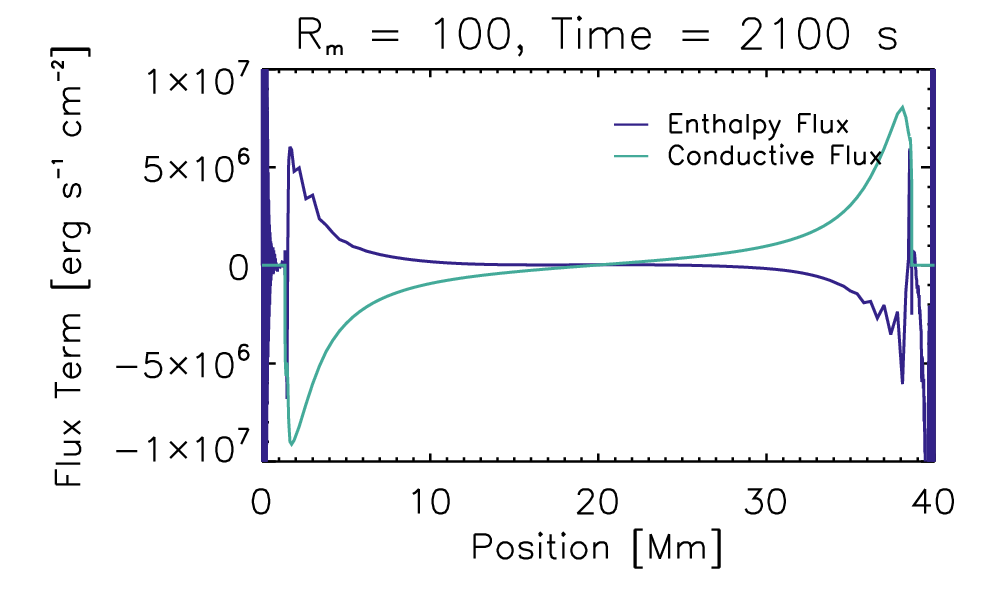}
    \includegraphics[width=0.31\textwidth]{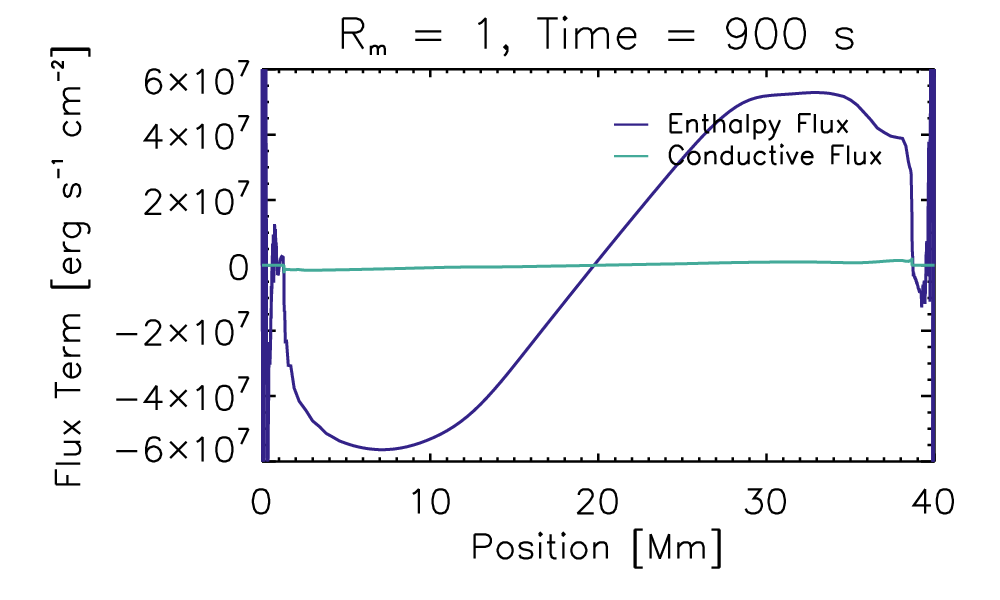}
    \includegraphics[width=0.31\textwidth]{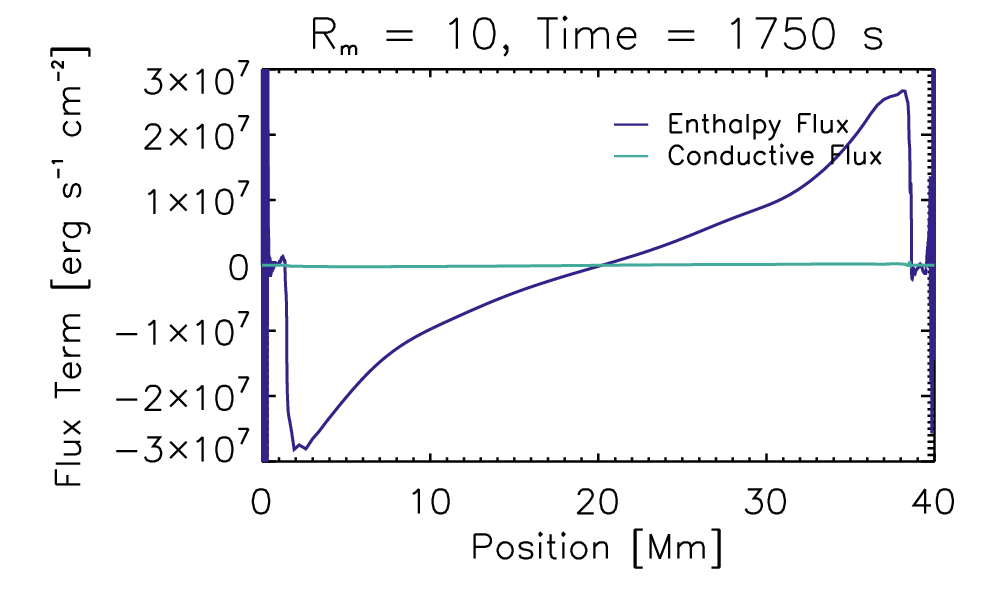}
    \includegraphics[width=0.31\textwidth]{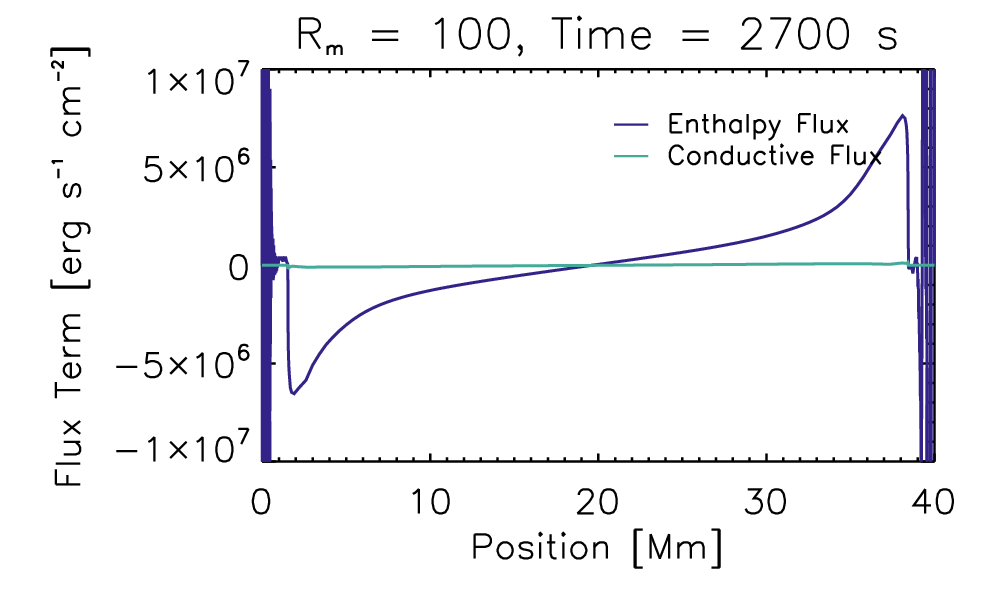}
    \caption{The evolution of the enthalpy flux, $v(P+E)$, and conductive heat flux, $\kappa_{0} T^{5/2} \partial_{s} T$, for the same simulations and times as in Figure \ref{fig:energy_terms}.  In the expanding cases, during early radiative cooling phase shown in the second row, the two terms approximately balance each other in the corona, as in Equation \ref{eqn:parity}, which causes the sustained upflow.  This parity breaks as radiation grows stronger with time.  The uniform loop, in contrast, does not show parity between the two terms.
 \label{fig:flux_terms}}
\end{figure*}

Finally, we also briefly show the evolution of the enthalpy flux, $v(P+E)$, and conductive heat flux, $\kappa_{0} T^{5/2} \partial_{s} T$, for the same times and simulations in Figure \ref{fig:flux_terms}.  In the two cases with expansion, when there is a sustained upflow (second row), the two terms are approximately equal and opposite, confirming Equation \ref{eqn:parity}.  The parity is slowly broken over the course of the cooling phase, and disappears entirely during catastrophic collapse.  Without expansion, that parity does not generally exist, and so the duration of upflows is tightly tied to the duration of heat deposition (see also \citealt{reep2018}).

\subsection{Cooling from a hot, static loop}

To separate out the effects of heating from cooling, we now present a case where we initialize the model with a hot loop that we simply allow to cool.  In the case of an impulsively heated loop, a large rise in the pressure due to the heating drives evaporation of material into the corona, and this increased pressure in the chromosphere may be partially responsible for the apparent sustained upflow.  In this case, we simply assume that a loop has been pre-heated, perhaps steadily, to reach a static equilibrium, and then we let it cool in order to see whether it develops similar upflows.

\begin{figure*}
    \centering
    \includegraphics[width=0.49\textwidth]{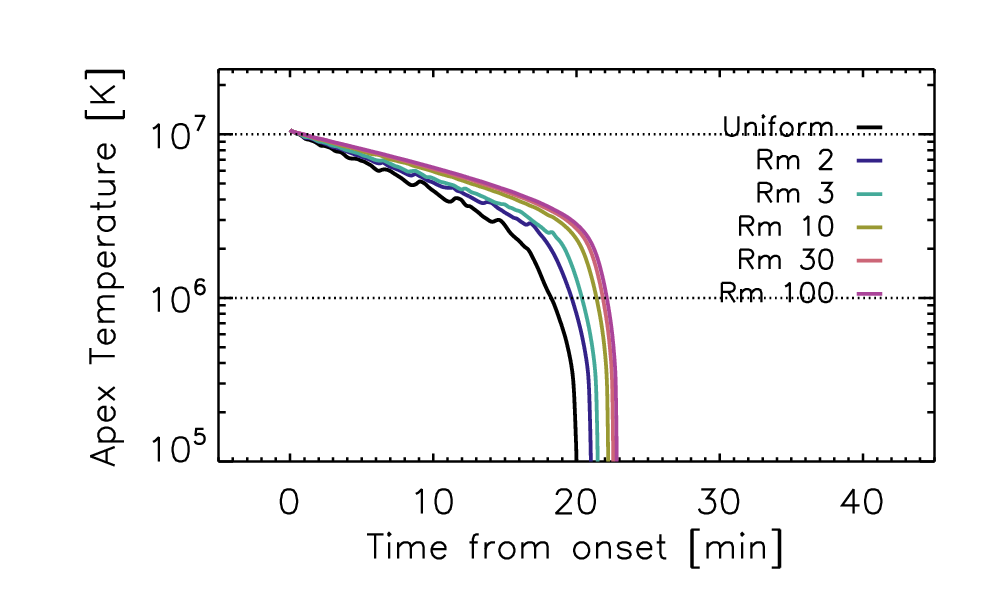}
    \includegraphics[width=0.49\textwidth]{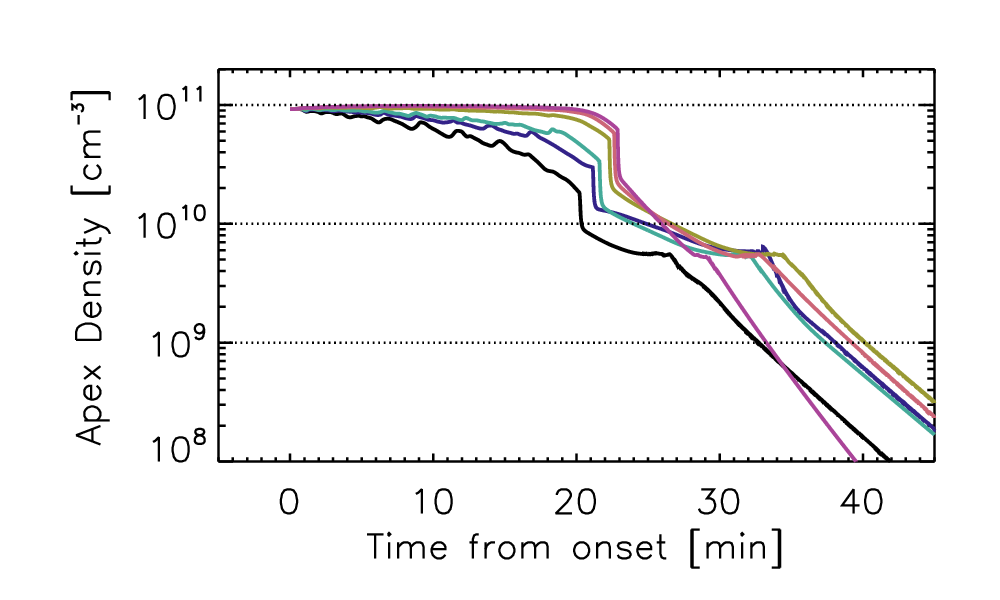}
    \includegraphics[width=0.49\textwidth]{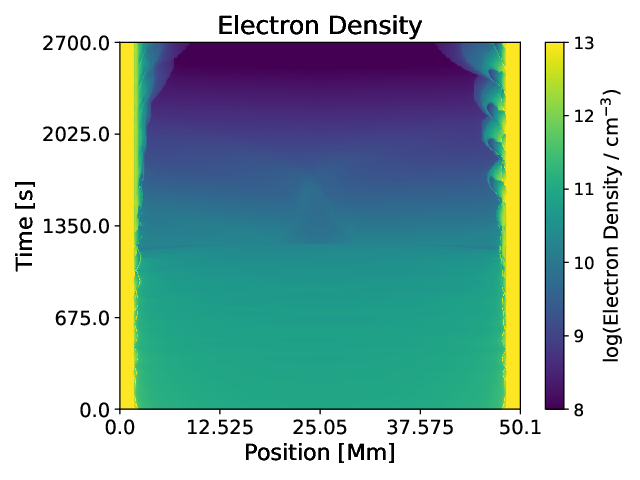}
    \includegraphics[width=0.49\textwidth]{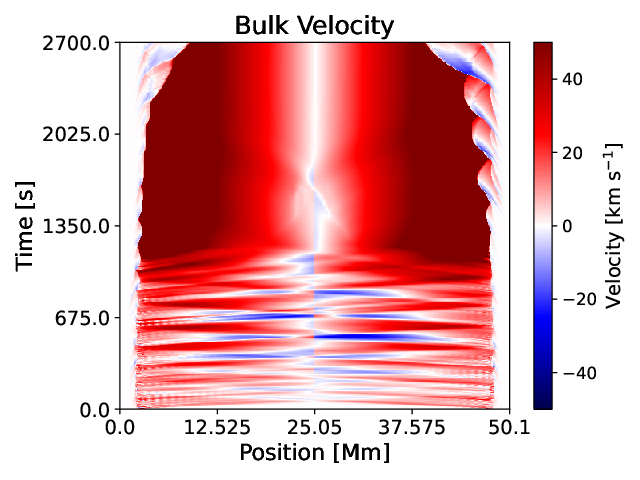}
    \includegraphics[width=0.49\textwidth]{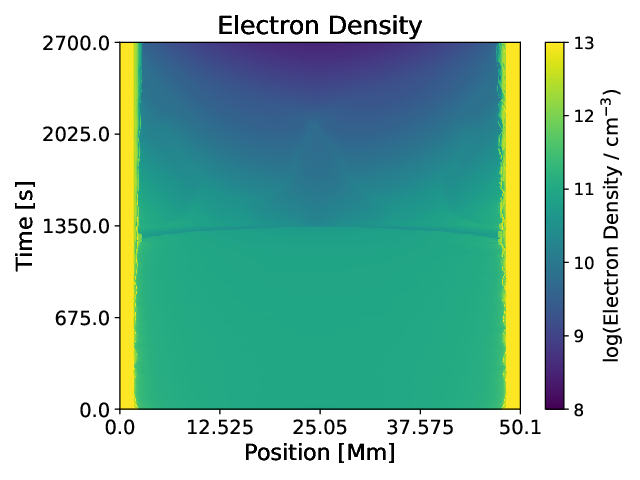}
    \includegraphics[width=0.49\textwidth]{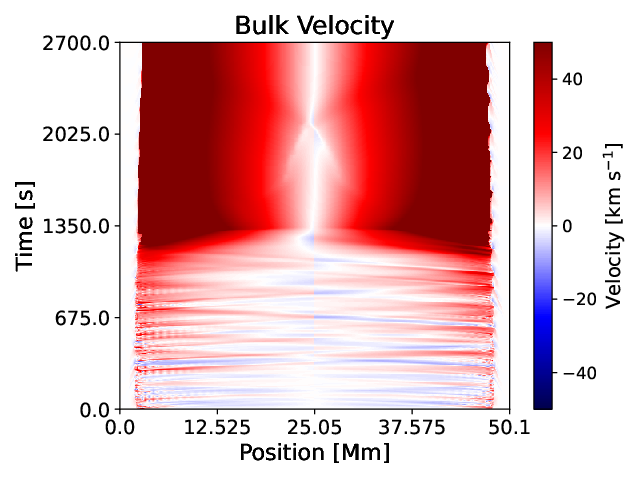}
    \caption{A comparison of pre-heated hot loops with various expansion factors that are simply allowed to cool.  Similar to Figure \ref{fig:heated}, where the middle row shows the uniform loop and the bottom row shows the case with expansion of 10.  Since there is no heating event in the simulation, there is no resultant evaporation, and the loops take similar times to cool without any long-lasting upflow.  \label{fig:cooling}}
\end{figure*}

Figure \ref{fig:cooling} shows the behavior of six such loops.  We examine loops of 50.1 Mm length, and vary the cross-sectional area expansion by a factor of 1, 2, 3, 10, 30, and 100, assuming a $\sin^{2}$ profile, as previously discussed in Section \ref{subsec:impulsive}.  The loops are assumed to be pre-heated, so that the initial coronal profile is hot $\approx 10$ MK and dense $\approx 10^{11}$ cm$^{-3}$.  We do not use any heating events in these simulations, nor coronal background heating, and simply allow the loops to cool.  The top row of Figure \ref{fig:cooling} shows the apex temperatures (left) and densities (right) as a function of time as the loops cool.  The uniform loop cools slightly faster, but all have similar cooling times.  Before the onset of catastrophic cooling, at about 20 minutes, the expanded loops only drain a minimal amount.  The other rows show a comparison of the spatiotemporal evolution of the electron densities (left) and bulk velocities (right) for the uniform loop (center row) and loop with expansion of 10 (bottom).  We can see that there are no significant differences in the evolution between the different cases, except perhaps that sound waves are more noticeable in the uniform loop (as noted also by \citealt{reep2022}).  

Unlike prior cases where we have heated the loops, the evolution of loops with and without expansion is quite similar.  There is no long-lasting induced upflow that we see with strong impulsive heating, showing that the initial pressure increase from a heating event is a necessary ingredient.  The cooling time becomes only slightly longer with expansion, and there is less draining before the catastrophic collapse.  For a hot, dense loop in equilibrium, conduction smooths the density and temperature profile across the length of the corona.  As a result, the velocities are slowed with larger expansion.  That is, in steady state, the mass flow rate

\begin{align}
    \partial_{s}(\rho A v) &= 0 \nonumber \\  
                            &\approx \rho\ \partial_{s}(A v)
\end{align}

\noindent so $A \times v$ is constant.  As the area grows with height, the velocity slows.  As a result, the cases with larger expansions drain more slowly.

This experiment shows that the drastically lengthened cooling times we find with loop expansion is due to the extra energy from the induced upflow carried into the corona, effectively reenergizing the corona.  Furthermore, the induced flow in expanded loops is a direct result of the initial impulsive heating burst, and the increased thermal conduction merely maintains that flow.

\subsection{Synthetic Lines}
\label{subsec:lines}

\citet{graham2015} found a consistent pattern of evolution of Doppler shifts all along the ribbon in a large flare observed with the Interface Region Imaging Spectrograph (IRIS; \citealt{depontieu2014}).  The \ion{Fe}{21} emission forms strongly blue-shifted, around 250-300 km s$^{-1}$, falling to zero velocity over the course of approximately 8-10 min.  This behavior is difficult to reconcile with short heating durations in a uniform area loop, which are commonly assumed to be less than 20 s.  On the other hand, chromospheric \ion{Mg}{2} emission was found to form strongly red-shifted, to perhaps 100 km s$^{-1}$, quickly decaying over the course of 1 to 2 min, in line with predictions of chromospheric condensations by \citet{fisher1989}.  Similar results were found by \textit{e.g.} \citet{polito2015,polito2016}.  Observations of flows with the Extreme Ultraviolet Imaging Spectrometer (EIS; \citealt{culhane2007}) onboard Hinode in lines forming across a wide temperature range, additionally, show that the magnitude of the velocity depends on the formation temperature \citep{milligan2009}.  It was also noticed, however, that TR lines like \ion{Si}{4} 1402 \AA\ often form red-shifted and remain red-shifted at around 10-20 km s$^{-1}$ for significantly longer than 10 min \citep{warren2016}, which is difficult to reconcile with a single loop model \citep{reep2016}.  \citet{reep2018} tried to understand this through forward modeling of \ion{Fe}{21} and \ion{Si}{4} Doppler shifts, finding that the approximately steady velocity in \ion{Si}{4} requires a multi-threaded solution with many loops rooted in one IRIS pixel ($\approx 350$ km), and that the duration of blue-shifts in \ion{Fe}{21} suggested that heating durations must be on the order of 100 s on average, much longer than expected.  This prior modeling work, however, assumed that the loops had uniform area.

\begin{figure*}
    \begin{minipage}[b]{0.88\linewidth}
    \centering
    \includegraphics[width=0.31\textwidth]{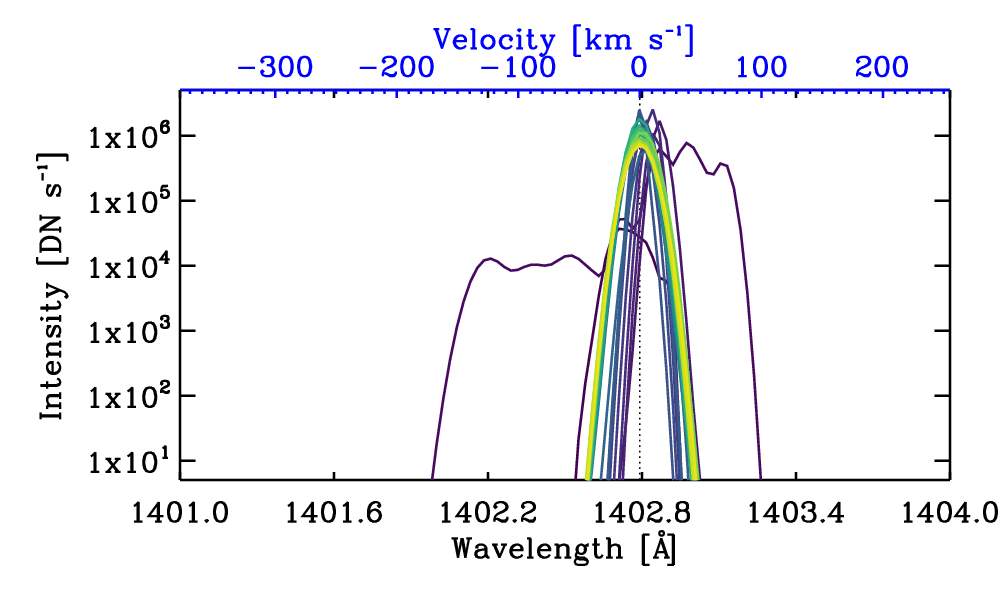}
    \includegraphics[width=0.31\textwidth]{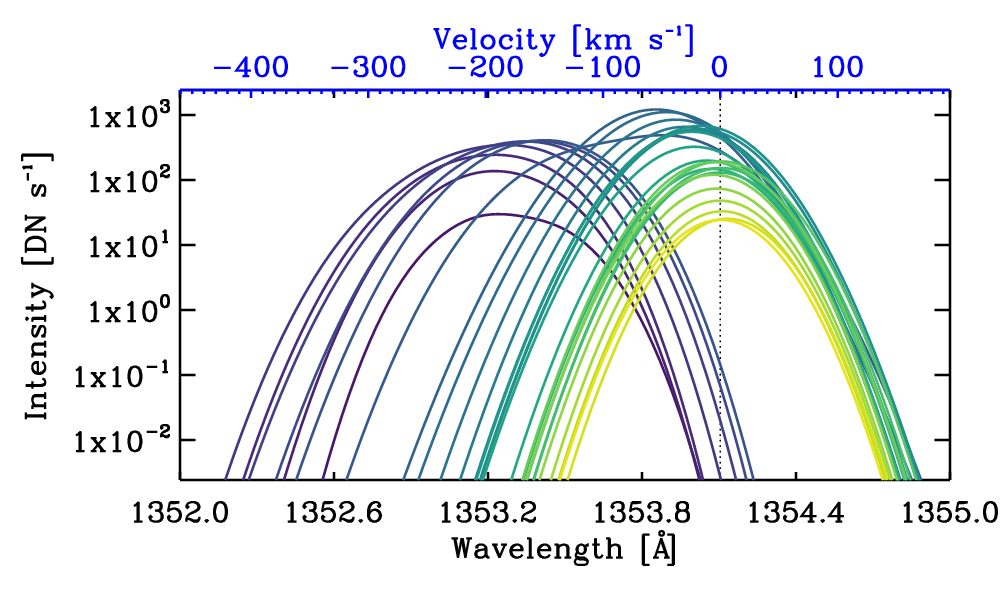}
    \includegraphics[width=0.31\textwidth]{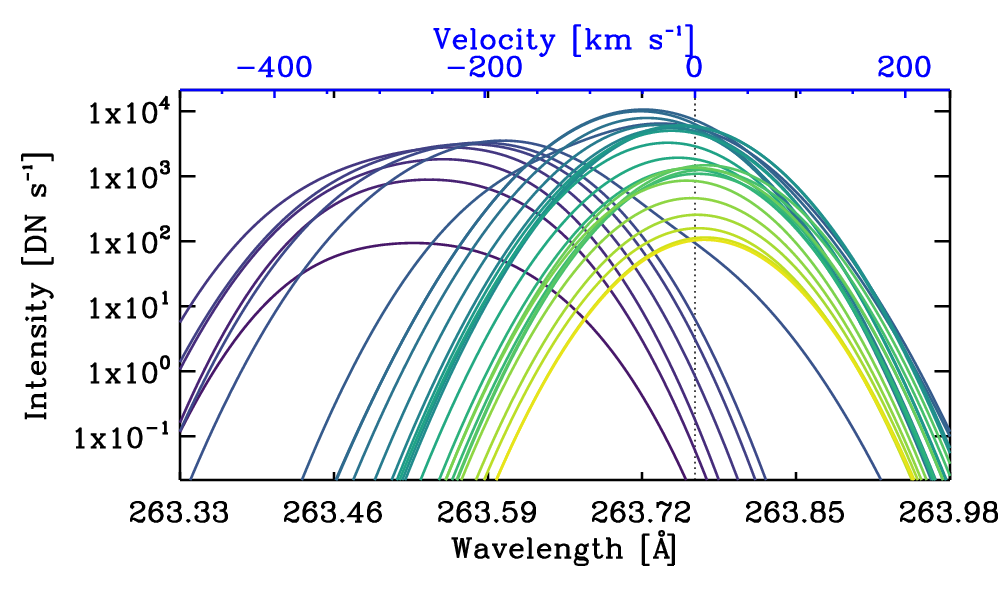}
    \includegraphics[width=0.31\textwidth]{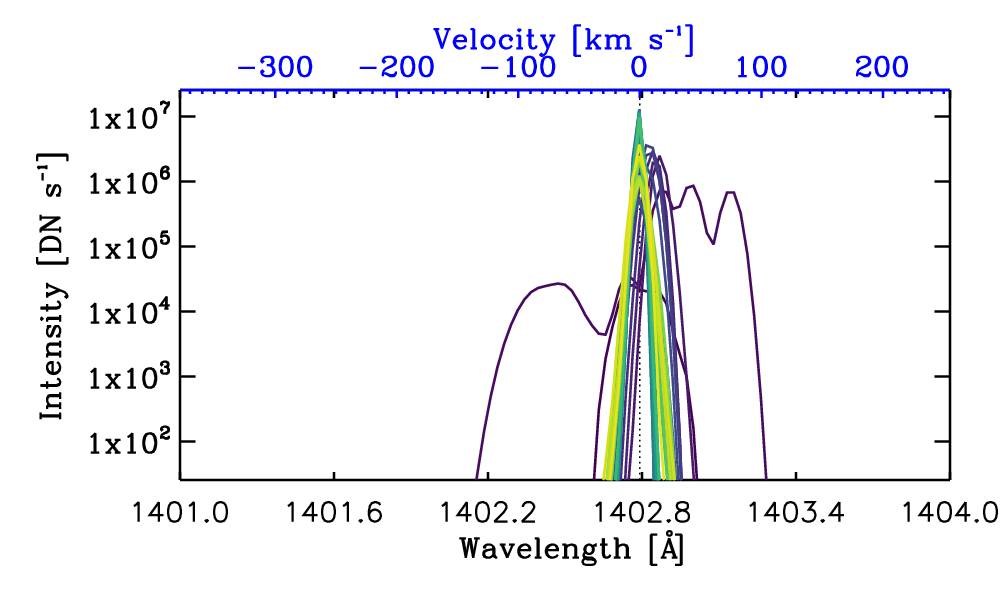}
    \includegraphics[width=0.31\textwidth]{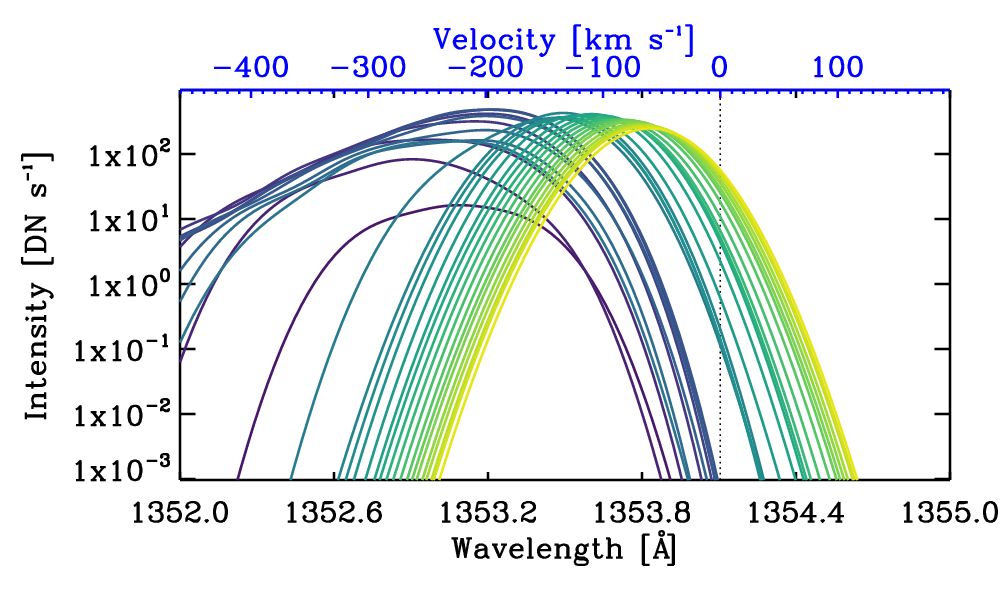}
    \includegraphics[width=0.31\textwidth]{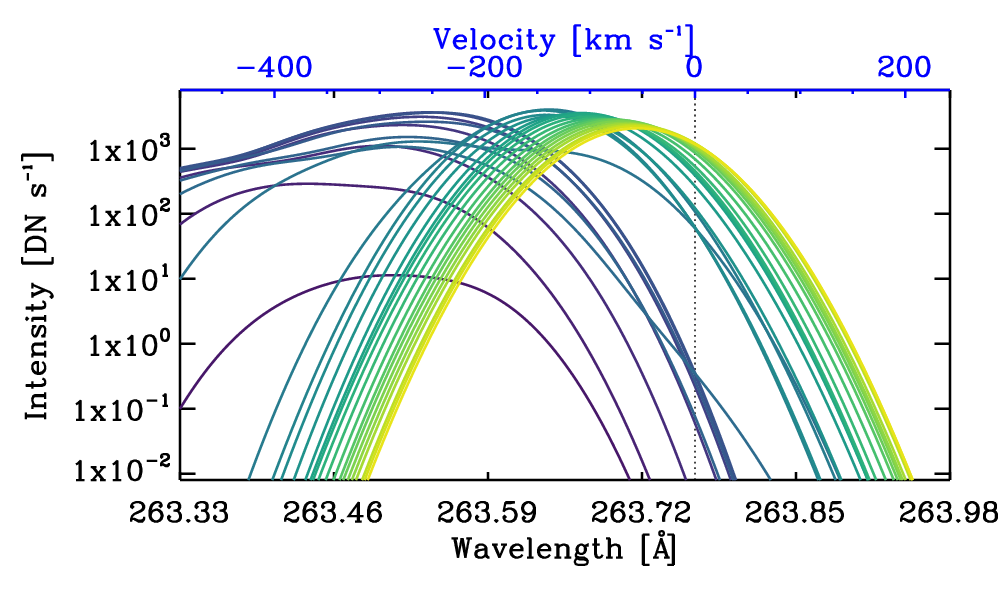}
    \includegraphics[width=0.31\textwidth]{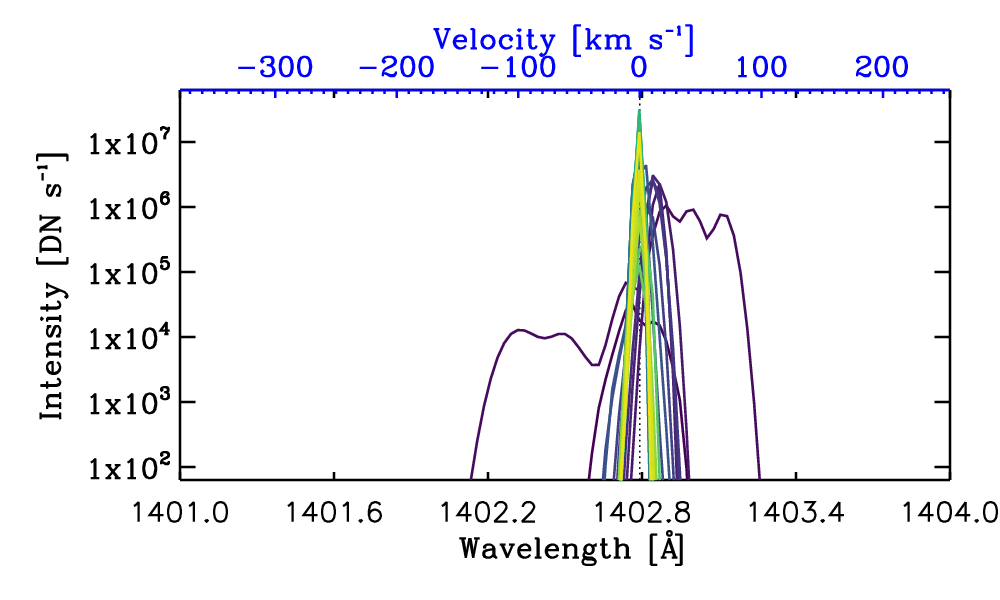}
    \includegraphics[width=0.31\textwidth]{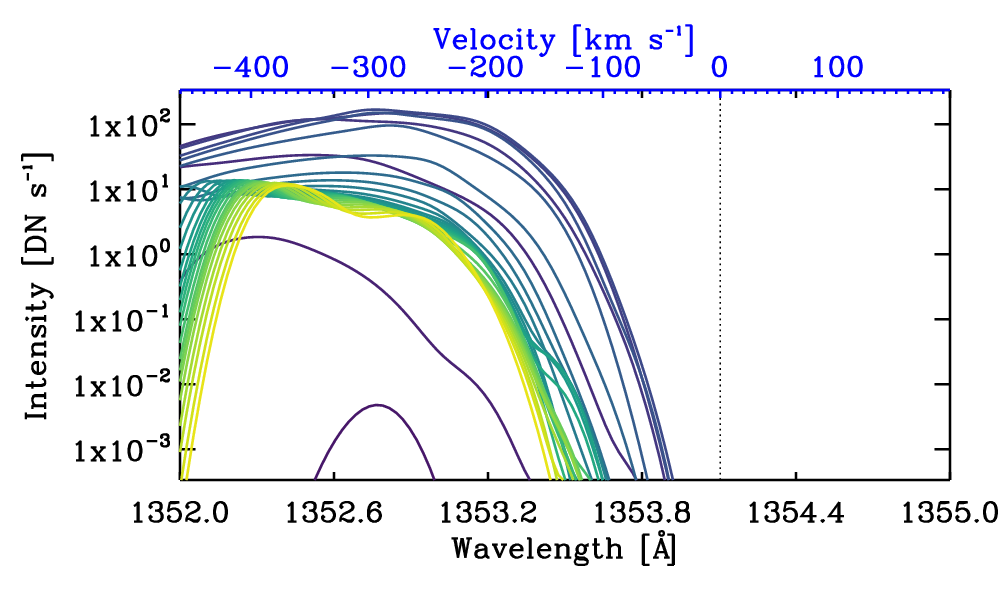}
    \includegraphics[width=0.31\textwidth]{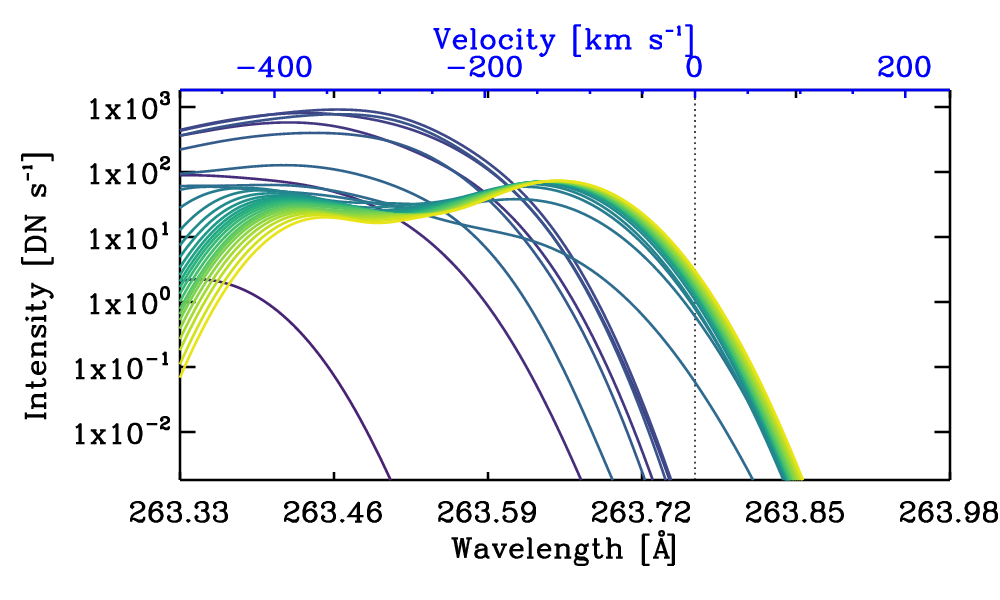}
    \end{minipage}
    \hfill
    \begin{minipage}[b]{0.1\linewidth}
     \includegraphics[width=\textwidth]{evap_colorbar.png}   
    \end{minipage}
    \caption{Synthetic lines as might be seen by IRIS and Hinode/EIS at a pixel at the footpoints of the loops, at a 10 second cadence for the first 300 s after the onset of heating for the simulations in Figure \ref{fig:strong_evap}.  The columns show IRIS \ion{Si}{4} 1402.77 \AA\ ($\log{T} = 4.85$), IRIS \ion{Fe}{21} 1354.08 \AA\ ($\log{T} = 7.05$), and EIS \ion{Fe}{23} 263.766 \AA\ ($\log{T} = 7.15$) respectively.  The rows show loops with uniform area, expansion of 10, and expansion of 100, respectively.  The cool \ion{Si}{4} emission does not significantly depend on the expansion, which occurs primarily in the corona.  Because of the induced upflows, however, the cases with expansion remain blueshifted for significantly longer than the heating duration, only gradually slowing with time.
 \label{fig:lines}}
\end{figure*}

The induced upflow seen in the simulations of loops with expanding area might provide an alternate explanation for the behavior.  That is, the expansion can provide a long-lasting upflow that could explain the \ion{Fe}{21} emission seen by \citet{graham2015} and others.  We have used the simulations in Figure \ref{fig:strong_evap} to forward model the behavior of \ion{Si}{4} and \ion{Fe}{21} as might be observed by IRIS and \ion{Fe}{23} as might be observed by Hinode/EIS (the weaker heating case in Figure \ref{fig:heated} did not produce significant \ion{Fe}{21} or \ion{Fe}{23} emission).  We use the forward modeling methodology of \citet{bradshaw2011}, which primarily assumes that the loop is at disk center, and we focus on the first pixel of the loop, which encompasses the footpoint emission.  We use the instrumental response of IRIS and EIS to synthesize the emission in detector units, and have assumed ionization equilibrium.  We show the resultant line profiles in Figure \ref{fig:lines}.

\begin{figure*}
    \centering
    \includegraphics[width=0.95\textwidth]{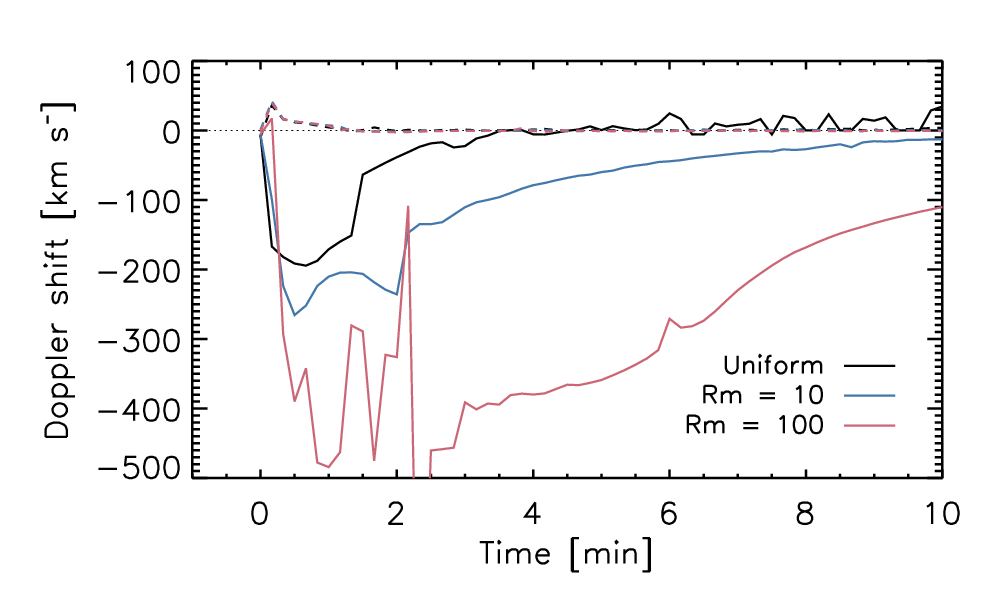}
    \caption{Fitted Doppler shifts of IRIS \ion{Fe}{21} 1354.08 \AA\ (solid) and IRIS \ion{Si}{4} 1402.77 (dashed) for the three cases in Figure \ref{fig:lines}.  The \ion{Si}{4} lines are briefly redshifted, then remain at rest.  The \ion{Fe}{21} lines are strongly blueshifted during the evaporation phase, and, with area expansion, remain blueshifted for long periods of time.
 \label{fig:doppler}}
\end{figure*}

The long-lasting upflow causes a long-lasting blueshift, in proportion to the total expansion factor.  We have fit the spectral line profiles with a single Gaussian component to measure their Doppler shifts.  The lines can have asymmetries and multiple components, however, so this is only meant to demonstrate the trends.  Figure \ref{fig:doppler} shows the Doppler shifts for \ion{Si}{4} 1402.77 \AA\ (dashed) and \ion{Fe}{21} 1354.08 \AA\ (solid).  For loops with or without expansion, the \ion{Si}{4} emission is briefly redshifted to around 20-30 km s$^{-1}$, and then returns to rest.  For \ion{Fe}{21}, however, the Doppler shifts depend strongly on the area expansion.  In the uniform case, the line becomes strongly blueshifted during the initial evaporation phase, and then returns to rest shortly after heating ceases (see also \citealt{reep2018}).  With expansion, however, the induced upflow causes the line to remain blueshifted for longer periods of time, only gradually slowing with time.  The duration of this Doppler shift is in direct proportion to the duration of the induced upflow, which lasts through much of the cooling phase, until radiation grows strong enough to break the parity of Equation \ref{eqn:parity}.  Of course, the line intensity would fade long before the line returns to rest since the plasma is also cooling, so this might not be detectable in actual observations.  

It is apparent that the duration of blueshifted emission not only depends on the heating duration as found by \citet{reep2018}, but also upon the rate, magnitude, and location of area expansion.  Future in-depth work is required to fully understand the parameter space (the flow parameters depend on all of the heating parameters and the exact geometry), and determine what heating profiles and rates of area expansion produce Doppler shifts and intensities consistent with observations like \citet{graham2015}.  Furthermore, we can see that the \ion{Si}{4} emission quickly returns to rest, suggesting that a multithreaded model is still required to explain the long-duration redshifts often seen in that line \citep{warren2016,reep2016}.

\section{Discussion}
\label{sec:disc}

A non-constant cross-sectional area significantly modifies the behavior of flows, which is well-known from fluid dynamics, but insufficiently considered in coronal loop dynamics.  The vast majority of coronal loop simulations to date have assumed a constant cross-section, which is guided by the observation that loop widths do not appear to vary along their length \citep{klimchuk2000}.  \citet{peter2012} suggested that this observation is because the temperature transverse to the field direction is not constant, and so a loop appears to have constant width when seen in spectral lines that form in a narrow temperature range.  In any case, assuming that a loop's area does vary spatially, the impact on flows modifies the cooling of, evaporation into, and draining from the loop.

In the simple case of a steady, subsonic flow, we find that the velocity change between chromosphere and corona is given by the relation $\frac{v_{2}}{v_{1}} = \frac{T_{2}}{T_{1}} \frac{A_{1}}{A_{2}}$.  Flows are relatively steady with strong evaporation, even steadier during catastrophic collapse of the loop, and become steadier still with larger expansion rates.  If expansion occurs all across the coronal loop, then there should be a noticeable change in velocity with height, seen for example in Figure 2 of \citet{reep2022}.  If expansion is confined near the TR, then the velocity should jump there, which is shown by Figure 5 of \citet{reep2022}, and similar behavior can be also seen with the localized expansions in \citet{mason2023}.  As a result, the cooling times are not as drastically lengthened since less energy is carried back into the corona.

After the onset of an impulsive heating event, evaporation drives material into the corona.  The initial flows do not occur in steady state, but tend towards steady state over time.  Furthermore, both a larger area expansion and a larger heating rate make the flows steadier.  With a large area expansion, additionally, because there is an approximate parity between the conductive and enthalpy fluxes (Equation \ref{eqn:parity} and Figure \ref{fig:flux_terms}), the energy carried out of the corona by conduction is resupplied by a sustained upflow.  This induced upflow lasts through the radiative cooling phase, increasing the coronal density marginally while the coronal temperature slowly decreases.  It ceases once the temperature has cooled enough that thermal conduction is too weak to power the radiative losses at the base of the corona.  Because the flow is supplying energy to the corona, the total cooling time of the loop increases.  At the onset of catastrophic collapse of the loop, however, the corona begins to rapidly drain, which occurs in steady state.

The induced upflows also cause an observational effect: there is a noticeable long-lasting blueshift that should be apparent in hot lines like \ion{Fe}{21} 1354 \AA.  Cooler lines forming in the TR are not drastically affected, showing similar behavior with and without area expansion.  When there is a large area expansion, the blueshifts in hotter lines take longer to return to rest, and that duration appears tied to the magnitude of expansion.  This could potentially be used to diagnose the area expansion (Figure \ref{fig:doppler}), but a significantly larger parameter study is required to compare with observations considering that many parameters can impact the properties of the evaporation (heating rate, heating duration, depth of heating, area expansion rate and location, etc.)  Further work is required to fully understand the diagnostic potential, particularly with high cadence IRIS \ion{Fe}{21} spectra.

\nopagebreak

\leavevmode \newline

\acknowledgments  
JWR, RBS, and KJK were supported by the Office of Naval Research 6.1 Support Program. Additionally, JWR is supported by a NASA Heliophysics Supporting Research grant under ROSES NNH19ZDA001N; KJK is supported by NASA's ISFM program ``Heating of the Magnetically Closed Corona'' (PI: Jim Klimchuk); and RBS is supported by a NASA Heliophysics Supporting Research grant under ROSES NNH20ZDA001N ``Investigating the Influence of Coronal Magnetic Geometry on the Acceleration of the Solar Wind'' (Science PI: RBS). This research benefited from discussions held at the International Space Science Institute in Bern, Switzerland, led by Drs. Vanessa Polito and Graham Kerr.

\bibliography{apj}
\bibliographystyle{aasjournal}

\end{document}